\pdfoutput=1
\documentclass{article}

\usepackage{natbib}

\pdfoutput=1

\usepackage[margin=1in]{geometry}

\usepackage{epsfig,amsmath, amsfonts,amsthm,amssymb}
\usepackage{dsfont}
\usepackage{esvect}
\usepackage{bm}

\usepackage[english]{babel}

\usepackage{stackrel}
\usepackage{verbatim}
\usepackage{setspace}
\usepackage{mathrsfs}
\usepackage{bm}
\usepackage{color}
\usepackage{epstopdf}
\usepackage{graphicx}
\usepackage{graphics}
\usepackage{multirow}
\usepackage[official]{eurosym}

\DeclareMathOperator*{\argmin}{arg\,min}

\usepackage[hyphens,spaces,obeyspaces]{url}

\usepackage{colortbl}

\newtheorem{thm}{Theorem}[section]
\newtheorem{cor}[thm]{Corollary}
\newtheorem{lem}[thm]{Lemma}

\newtheorem{prop}[thm]{Proposition}
\theoremstyle{definition}

\newtheorem{example}{Example}

\newtheorem{rem}[thm]{Remark}
\numberwithin{equation}{section}
\newcommand{\be}{\begin{equation}}
\newcommand{\ee}{\end{equation}}
\newcommand{\bq}{\begin{eqnarray}}
\newcommand{\eq}{\end{eqnarray}}

\newcommand{\half}{\frac{1}{2}}

\def\calF{{\mathcal F}}

\def\calG{{\mathcal G}}

\def\bbrZ{{{\mathbb R}_0}}

\def\bbr{{\mathbb R}}
\def\bbe{{\mathbb E}}
\def\bbp{{\mathbb P}}

\def\bbn{{\mathbb N}}

\def\b{{b}}

\definecolor{Red}{rgb}{1.00, 0.00, 0.00}
\newcommand{\Red}{\color{Red}}
\definecolor{DRed}{rgb}{0.7, 0.3, 0.00}

\definecolor{Green}{rgb}{0.2, 0.5, 0.2}

\definecolor{Blue}{rgb}{0.00, 0.00, 1.00}

\definecolor{PaleGrey}{rgb}{.6, .6, .6}

\usepackage{listings}
\definecolor{mygreen}{RGB}{28,172,0} 
\definecolor{mylilas}{RGB}{170,55,241}

\date{}

\title{Personalized Robo-Advising:\\ Enhancing Investment through Client Interaction\footnote{We are grateful for constructive comments from Peter Carr, Anthony Ledford, Charles-Albert Lehalle, Alberto Rossi, and seminar participants at Vanguard, the Robo-Advising Day at Georgetown University's Center for Financial Markets and Policy, the University of Southern California, Boston University, the New England Statistics Symposium, the CFS Workshop on AI/ML in Financial Services, the Oxford-Man Institute, the NUS Quantitative Finance Series, the Brooklyn Quant Experience Lecture Series, the ACPR Workshop on Robo-Advising, the New Ideas in Quantitative Finance Workshop at Stony Brook University, the 2019 SIAM-FM Annual Meeting, the 2019 ICIAM, and the 2020 INFORMS Annual Meeting.}}

\author{Agostino Capponi\thanks{Department of Industrial Engineering and Operations Research,
		Columbia University, New York, NY 10027, USA,
		\texttt{ac3827@columbia.edu.}} \quad
	Sveinn \'Olafsson\thanks{Department of Industrial Engineering and Operations Research,
		Columbia University, New York, NY 10027, USA,
		\texttt{so2570@columbia.edu}} \quad
	Thaleia Zariphopoulou\thanks{Departments of Mathematics and IROM, The University of Texas at Austin, and the Oxford-Man Institute, University of Oxford, \texttt{zariphop@math.utexas.edu}}}

\begin{document}
	
	\maketitle
	
	\doublespacing
 
\begin{abstract}
	Automated investment managers, or robo-advisors, have emerged as an alternative to traditional financial advisors. The viability of robo-advisors crucially depends on their ability to offer personalized financial advice. {We introduce a novel framework, in which a robo-advisor interacts with a client to solve an adaptive mean-variance portfolio optimization problem.} The risk-return tradeoff adapts to the client's risk profile, which depends on idiosyncratic characteristics, market returns, and economic conditions. We show that the optimal investment strategy includes both myopic and intertemporal hedging terms which are impacted by the dynamics of the client's risk profile. We characterize the optimal portfolio personalization via a tradeoff faced by the robo-advisor between receiving client information in a timely manner and mitigating behavioral biases in the risk profile communicated by the client.
	We {argue that} the optimal portfolio's Sharpe ratio and return distribution {improve if} the robo-advisor counters the client's tendency to reduce 
	market exposure during economic contractions when the market risk-return tradeoff is {more favorable}. 
\end{abstract}

\section{Introduction}\label{sec:intro}

Automated investment managers, commonly referred to as robo-advisors, have gained widespread popularity in recent years. The value of assets under management
by robo-advisors is the highest in the United States, exceeding \$600 billion in 2019. Major robo-advising firms include Vanguard Personal Advisor Services, with about \$140 billion of assets under management, Schwab Intelligent Portfolios (\$40bn), Wealthfront (\$20bn), and Betterment (\$18bn). 
Robo-advisors are also on the rise in other parts of the world, managing over \$100 billion in Europe, and exhibiting rapid growth in Asia where the assets under management exceed \$75 billion solely in China (\cite{Statista}). 

The first robo-advisors were launched in the wake of the 2008 financial crisis and the ensuing loss of trust in established financial services institutions. 
Examples of pioneering robo-advising firms are Betterment and Wealthfront, which began offering services formerly considered exclusive to the general public, including individuals who did not meet the minimum investment levels of traditional financial advisors.  
In the years that followed, industry incumbents - such as Vanguard and Charles Schwab - followed suit and began offering their own robo-advising services, taking advantage of their existing customer bases to quickly gain a large market share.
With the focus of most robo-advisors being on passive long-term portfolio management, the rise of robo-advising since the financial crisis has also been compounded by the seismic shift towards passive investing and low cost exchange-traded funds.\footnote{See \url{https://www.cnbc.com/2018/09/14/the-trillion-dollar-etf-boom-triggered-by-the-financial-crisis.html}}

Robo-advising is a term that encompasses various forms of digital financial advice for investment management and trading. 
A recent survey paper by \cite{Acunto2} distinguishes between digital tools and services that aid in active, short-term trading, with the client actively involved in the strategy implementation, 
and robo-advisors {that focus} on long-term passive investing where the level of delegation is higher. 
{We develop a framework for automated investment management which belongs to the latter category, where the robo-advisor uses quantitative methods and algorithms to {manage} the client's portfolio. We explore how the robo-advisor can use features of the client's risk profile to determine a frequency of interaction with the client that ensures a high level of portfolio personalization.\footnote{Robo-advisors are fiduciaries under the Investment Advisers Act of 1940, and as such are subject to the duty of acting in the client's best interest.} We also analyze the dilemma faced by the robo-advisor in either catering to the client's wishes, i.e., investing in accordance with the client's risk profile, or going against the client's wishes in order to seek better investment performance.

	In our framework, the  robo-advisor interacts repeatedly with the client to learn about changes in her risk profile. This risk profile is characterized by a risk aversion process, which captures three fundamental aspects of individual investor behavior.  
	First, the client's risk aversion is sensitive to the passage of time, consistent with empirical research that has identified a positive and potentially nonlinear trend in risk aversion as a function of age (see, e.g., \cite{Brooks} and \cite{Hallahan}). This type of time-dependence has also been used in the construction of Target Date Funds, which are widely used retirement portfolios that reduce equity risk as the client gets closer to retirement.\footnote{Target Date Funds are similar to the ``set-and-forget'' portfolios offered by the first generation of robo-advisors, {where changes in allocation over time are solely based} on the client's age. While TDFs have virtually stayed the same since their inception over a quarter century ago, the technology-driven personalized advice of robo-advisors has continued to evolve.} 
	Second, risk aversion is impacted by idiosyncratic shocks, such as a change in disposable income or an increase in the client's educational level or financial literacy (see, e.g, \cite{Hallahan} and \cite{Allgood}). These shocks in particular {capture changes in unobservable factors driving the client's risk aversion}, consistent with empirical studies that {have identified} a substantial variation in idiosyncratic risk preferences that are unexplained by consumer attributes and demographic characteristics (see, e.g., \cite{GuisoPaiella,Sahm}, and \cite{Venter}). Third, the client's risk aversion depends on realized market returns and prevailing economic conditions. Specifically, the countercyclical nature of risk aversion has been extensively documented in the literature (see \cite{Bucciol}, \cite{Cohn}, and \cite{Guiso}, among others, for recent empirical and experimental evidence).

	The robo-advisor adopts a multi-period mean-variance investment criterion with a finite investment horizon. It is worth noting that most robo-advising firms employ asset allocation algorithms based on mean-variance analysis, {citing, among other things,} its tractability and the intuitive interpretation of the risk-return tradeoff (see, for instance, \cite{Beketov}). While these robo-advising algorithms assume the risk-return tradeoff coefficient to be constant throughout the investment period, and statically reoptimize the portfolio in response to change in the client's risk profile, our framework enables the robo-advisor to adapt the portfolio strategy to {\it its own model} of the client's risk aversion process. 
	Such a model is constructed from two sources of information. First, the robo-advisor observes both realized market returns and changes in the state of the economy, and continuously updates its model to reflect this information. Second, at times of interaction with the client, the robo-advisor receives information {about the} idiosyncratic component of the client's risk aversion.

	{We show that the optimal stock market allocation consists of two components. The first component is akin to the standard single-period Markowitz strategy but also takes into account the expected return and variance of the optimal strategy throughout the investment {period}. The second component is  intertemporal hedging demand, which depends on the relation between the current market return and future portfolio returns. A novel feature of our model is that this relation is driven by the client's dynamic risk profile, in contrast to extant literature (e.g., \cite{Liu} and \cite{Basak}) where it depends {only} on the correlation between market returns and changes in a stochastic state variable.
		To understand why hedging demand arises in our model, consider the case where positive (negative) market returns have a tempering (inflating) effect on the client's risk aversion process. A positive market return thus leads to increased future investment and higher anticipated portfolio returns, whereas a negative return has the opposite effect. This positive relation between market returns and future portfolio returns amplifies the effect of the current allocation on the variance of terminal wealth. As a result, for a client whose risk aversion is negatively correlated with market returns, there is a negative hedging demand, i.e., the robo-advisor reduces her market exposure relative to a client whose risk aversion is not sensitive to 
		market returns.		

		We show the existence of a tradeoff between the rate of information acquisition from the client and the accuracy of the acquired information. On the one hand, if interaction does not occur at all times, the robo-advisor may not always have access to up-to-date information about the client's risk profile. On the other hand, information communicated to the robo-advisor may not be representative of the client's true risk aversion, as the client is subject to behavioral biases, such as trend-chasing.\footnote{See \cite{Kahneman} for a discussion of common behavioral biases and strategies to overcome them.} For example, if recent market returns have exceeded expectations, the client may feel overly exuberant and communicate a risk aversion value that is lower than her actual risk aversion. Vice versa, following a market underperformance, the client may feel overly pessimistic and exaggerate her risk aversion when communicating with the robo-advisor. 
		The suboptimality of high interaction frequencies then hinges on the fact that reducing the frequency of interaction mitigates the effect of client's biases. 
		The rationale is that over longer time periods, 
		fluctuations in realized market returns average out. 
		Such a behavioral pattern is consistent with the notion of myopic loss aversion introduced in \cite{Benartzi}, which refers to the empirical phenomenon that market underperformance has a stronger impact on risk aversion than market overperformance.

		We introduce a measure of portfolio personalization to analyze the aforementioned tradeoff. This measure is defined in terms of the difference between the client's actual risk aversion and the robo-advisor estimate of it. {The lower this measure, the closer the investment strategy implemented by the robo-advisor is to the strategy matching the client's risk profile.} 
		We show that for a fixed level of behavioral bias, quantified as the client's sensitivity to market return fluctuations, this measure is minimized by a unique interaction frequency.  
		Such a frequency allows the robo-advisor to strike a balance between obtaining information from the client in a timely manner and ensuring that the communicated information is not overly {biased} by recent market returns. Our result is supported by existing practices of robo-advising firms, which  encourage clients to refrain from making frequent changes to their risk profiles, and even limit their ability to do so.\footnote{Wealthfront provides the following guidelines: ``Our software limits our clients to one risk-score change per month. We encourage people who attempt more than three risk-score changes over the course of a year to try another investment manager.''} 
		We also {show} that the optimal interaction frequency is decreasing in the {amount of behavioral bias,  and increasing in the rate at which the idiosyncratic component of the client's risk aversion changes. 
			Therefore, a higher level of personalization is achieved for clients with limited behavioral biases and stable risk profiles. 

			We compare the investment performance of strategies that invest the same proportion of wealth in the stock market during periods of economic growth but differ in their market exposure during contractions. 
			We show analytically that the Sharpe ratio of such strategies is generally increasing in the proportion of wealth allocated to the stock market in a state of contraction, when the market risk-return tradeoff is more favorable.\footnote{Since the seminal work of \cite{Fama}, it has been extensively documented that the market risk-return tradeoff is countercyclical, i.e., higher at business cycle troughs than peaks.} Because the client typically has a countercyclical risk aversion process, she would lean towards reducing market exposure during economic contractions and, as a result, attain a lower Sharpe ratio from her investment. {The question that then arises is to what extent, if any, the robo-advisor should deviate from the strategy implied by the client's risk profile in order to improve the investment performance. We show that the Sharpe ratio of the optimal portfolio is concave with respect to the change in the market allocation when the economy moves to a state of contraction. This means that the Sharpe ratio drops more if the allocation is reduced, compared to how much it improves if the allocation is increased by the same amount. Our analysis thus suggests that a good middle ground for the robo-advisor is to rebalance the portfolio and maintain prespecified weights throughout the business cycle. 
				This investment pattern is also in line with the long-term investment principle of riding out business cycles, and further reinforced by the fact that, in practice, it is difficult to detect changes in economic regimes and thus estimate the prevailing market risk-return tradeoff.} 

			The remainder of the paper is organized as follows. In Section \ref{sec:litreview}, we briefly review {the} related literature. In Section \ref{SectionModel}, we introduce our robo-advising framework. In Section \ref{SectionOptimalAllocation}, we present the solution to the associated {optimal} investment problem. In Section \ref{SectionRegret}, we introduce and study performance metrics for the robo-advisor's investment strategy. 
			Section \ref{sec:conclusions} offers concluding remarks. 
			Appendix \ref{Appendix0} contains theoretical and computational properties of the optimal investment strategy. 
			Appendix \ref{AppendixA} and Appendix \ref{AppendixC} contain proofs {and auxiliary results} related to Sections \ref{SectionOptimalAllocation} and \ref{SectionRegret}, respectively. 
			
	
\section{Literature Review}\label{sec:litreview}

Our study contributes to the growing literature on robo-advising. \cite{Acunto2} describe the main components of robo-advising systems and propose a classification in terms of four main features: (i) portfolio personalization, (ii) client involvement, (iii) client discretion, and (iv) human interaction. We offer a quantitative framework that aligns with this classification and is consistent with that of the most prominent stand-alone robo-advising firms. We proceed to briefly describe each of the four features and how they fit into our framework. 

(i) \emph{Portfolio personalization} refers to the robo-advisor's ability to offer financial advice tailored to the client's needs. A common criticism of robo-advising is the lack of customization, with risk profiling based on information that is too limited.\footnote{Risk profiling can, in principle, be infinitely customizable. Barron's 2019 annual ranking of robo-advisors (available at \url{https://webreprints.djreprints.com/4642511400002.pdf)} reports the latest growth in robo-advising to be in cash management, with robo-advisors aiming to become everyday money managers of their clients, in addition to long-term investment planners. Robo-advising firms already offer cash management services such as direct deposits of paychecks, automatic bill payments, and FDIC-insured checking and savings accounts. This implies increased access to data which can in turn be used by robo-advisors to improve investment recommendations.}
In our framework, the robo-advisor achieves portfolio personalization by soliciting information from the client. We do not explicitly model the process of collecting information, which can be based both on online questionnaires and on collection of client's data from various sources (e.g. savings and spending behavior,  asset and liabilities).} 

(ii) \emph{Client involvement} refers to the client's participation in the design of the investment strategy. At one end of the spectrum, there are robo-advisors that require the client to approve every single trading decision. In this case, the client is actively involved and the term robo-\emph{advisor} is descriptive, because the client decides in what way to follow the \emph{advice} provided.
Our model lies at the other end of the spectrum, where the robo-advisor automatically \emph{manages} the portfolio on the client's behalf and for which the term robo-\emph{manager} is more descriptive.

(iii) \emph{Client discretion} {is} the client's ability to override the robo-advisor's recommendation. In our framework, the client's behavioral biases can be viewed as the client overriding the robo-advisor's recommendations. This is considered a low level of discretion. It is consistent with the operations of robo-advisors focusing on long-term investing, where the client is allowed to adjust the level of portfolio risk, but has no control over which parts of the investment portfolio are modified. 

(iv) \emph{Human interaction} refers to the degree of interaction {between the client and a human-advisor.} In our framework, there is no human contact 
which is also the case for automated robo-advisors that strive to minimize operating costs.

The above classification is qualitative in nature. The quantitative components of our framework are  constructed to {resemble the algorithmic principles} of actual robo-advisors. \cite{Beketov} provide 
an industry overview based on the analysis of 
over 200 robo-advisors globally. Their study shows that a large majority of robo-advisors use an asset allocation framework based on mean-variance analysis, with the asset universe consisting of low cost exchange-traded funds. 
They also show that risk profiling of clients is primarily {done using} online questionnaires.

\cite{Rossipeace} conduct an extensive survey to study the ``needs and wants'' of individuals when they hire financial advisors. 
{They show} that for traditionally-advised clients, algorithm aversion and the inability to interact with a human are the main obstacles for switching to robo-advising, while robo-advised clients do not have the same need for trust and  access to expert opinion. {Our model lends theoretical support to the notion that robo-advisors 
	{may be less suitable for} algorithmic-averse clients.} 
Namely, the robo-advisor can improve the portfolio performance by going against the wishes of the client, i.e., by investing in a way that may seem counterintuitive to {the client}. Doing so is more challenging for a robo-advisor than {for} a human-advisor, because market returns are random, and an algorithmic-averse client will be less forgiving to the robo-advisor in the event of adverse market returns.

From a methodological perspective, our work contributes to the literature on \emph{time-inconsistent} stochastic control (\cite{Bjork}). Other related works include \cite{LiNg} who solve a multi-period version of the classical Markowitz problem, and \cite{Basak} who solve a continuous-time version of the 
same problem. 
\cite{BjorkZhou} solve the dynamic mean-variance problem in continuous time, 
with the mean-variance utility function applied to {returns} {as in the classical} single-period mean-variance analysis. A recent study of \cite{KouRobo} further develops a dynamic mean-variance framework based on log-returns, with the aim of generating policies conforming with conventional investment wisdom. In all of these works, the risk-return tradeoff is assumed to be constant throughout the investment horizon. By contrast, in our model, the risk-return tradeoff coefficient is stochastic, with explicitly modeled dynamics that can be used to generate investment policies tailored to the client's risk profile. 

{Our paper is also related to the literature on portfolio optimization models where the security price dynamics are driven by stochastic factors.} 
Unlike \cite{Liu} and \cite{Basak}, intertemporal hedging demand does not arise because of dependence between the stochastic factor and market returns, which are assumed to be uncorrelated in our model. Rather, intertemporal hedging terms appear because of the correlation between market returns and future changes in the investor's risk aversion. Additionally, the stochastic factor itself impacts the client's risk aversion dynamics, which in turn determine the optimal allocation. It is widely agreed upon that risk aversion varies with the business cycle, and this characteristic is highly relevant for robo-advising where a significant challenge is to keep clients invested in the stock market during economic downturns.}\footnote{The seminal work of \cite{Campbell} provides an asset pricing framework that rationalizes the countercyclicality of both risk aversion and the market risk-return tradeoff; see, also, \cite{Lettau} for a survey of the literature and extensive empirical evidence.}

\section{Modeling Framework}\label{SectionModel}

Our robo-advising framework has four main components: (i) a regime switching model of market returns, (ii) a mechanism of interaction between the client and the robo-advisor, (iii) a {dynamic} model for the client's risk preferences, and (iv) an optimal investment criterion. We describe each of these components in {Sections} \ref{SectionModelRobo}-\ref{SectionInvestment}.

\subsection{Market and Wealth Dynamics}\label{SectionModelRobo}

The market consists of a risk-free money market account, $(B_n)_{n\geq 0}$, and a risky asset, $(S_n)_{n\geq 0}$, whose dynamics are described by the equations
\begin{align*}
B_{n+1}&=(1+r(Y_n))B_n,\qquad
S_{n+1}=(1+Z_{n+1}(Y_n))S_n,
\end{align*}
where $B_0=1$ and $S_0>0$ is a given constant. The risky asset is representative of the overall stock market, and for this reason we will refer to returns of the risky asset as market returns. 

The above dynamics are modulated by an observable state process, $(Y_n)_{n\geq 0}$, which captures macroeconomic conditions affecting interest rates and market returns {(cf.\ \cite{Hamilton})}. We assume {the economic state variable} $(Y_n)_{n\geq 0}$ to be a time-homogeneous Markov chain with transition matrix $P$, taking values in a finite set $\mathcal{Y}:=\{1,2,\dots,M\}$, for some $M\geq 1$. Conditioned on $Y_n=y\in\mathcal{Y}$, the risk-free interest rate $r(y)$ is constant, while the risky asset's return, $Z_{n+1}(y)$, 
admits a probability density function $f_{Z|y}$ {which depends only} on the current economic state $y$, and has mean $\mu(y)$ and variance $0<\sigma^2(y)<\infty$.  

For notational simplicity, we omit the dependence on the economic state. Hence, 
we use $Z_{n+1}$ in place of $Z_{n+1}(Y_n)$, and denote its state-dependent mean and variance by $\mu_{n+1}:=\mu(Y_n)$ and $\sigma_{n+1}^2:=\sigma^2(Y_n)$, respectively. 
Similarly, we use $r_{n+1}$ in place of $r(Y_n)$, and let $R_{n+1}:=1+r_{n+1}$. We denote by $\widetilde Z_{n+1}:=Z_{n+1}-r_{n+1}$ the excess return of the risky asset over the risk-free rate, {which has} state-dependent mean and variance given by $\tilde\mu_{n+1}:=\mu_{n+1}-r_{n+1}$ and $\sigma^2_{n+1}$, respectively. 

{We denote by $X_n$ the wealth of the client at time $n$}, allocated between the risky asset and the money market account, and use $\pi_n$ to denote the amount invested in the risky asset. For a {given} self-financing trading strategy $\pi:=(\pi_n)_{n\geq 0}$, 
the wealth process $(X^{\pi}_n)_{n\geq 0}$ follows the dynamics
\begin{align}\label{dX}
X_{n+1}^{\pi} = 
R_{n+1}X_n^{\pi} + \widetilde Z_{n+1}\pi_n,\qquad X_0^{\pi}=x_0.
\end{align}

{The initial wealth, $x_0\in\bbr^+$, and the} initial state of the economy, {$Y_0\in\mathcal{Y}$}, are assumed to be non-random. The random variables $(Y_n)_{n\geq 1}$ and $(Z_n)_{n\geq 0}$ are defined on a probability space $(\Omega,\calF,\bbp)$,
which additionally supports a sequence $(\epsilon_n)_{n\geq 1}$ of independent real-valued random variables. 
This source of randomness captures idiosyncratic changes to the client's risk preferences. Moreover, $(\epsilon_n)_{n\geq 1}$ is independent of $(Y_n)_{n\geq 0}$ and $(Z_n)_{n\geq 1}$. {We use $(\calF_n)_{n\geq 0}$ to denote the filtration} generated by the three stochastic processes in our model:
\begin{align}\label{calF}
\calF_n := \sigma(Y_{(n)},Z_{(n)},\epsilon_{(n)}),
\end{align}
where $Y_{(n)}:=(Y_0,\dots,Y_n)$, $Z_{(n)}:=(Z_1,\dots,Z_{n})$, and $\epsilon_{(n)}:=(\epsilon_1,\dots,\epsilon_n)$. We will use analogous notation to denote the paths of other stochastic processes throughout the paper.

\begin{rem}\label{S0neg}
	In our discrete-time model, the price of the risky asset is not restricted from becoming negative. However, at any given time the optimal investment strategy presented in Section  \ref{SectionOptimalAllocation} does not depend on the {current} price of the risky asset, but only {on} the first two moments of its returns distribution. Hence, for the optimal investment problem {to be well-defined,} it is sufficient to assume the existence of {a} returns process $(Z_n)_{n\geq 1}$ with a finite variance as described above. 
	\hfill\qed
\end{rem}

\subsection{Interaction between Client and  Robo-Advisor}\label{SectionModelInteraction}

A key component of the proposed framework is the interaction between the client and the robo-advisor. Following an initial interaction at the beginning of the investment process, the client and the robo-advisor interact repeatedly throughout the investment period. At each interaction time, the client translates the information communicated by the client into a numerical value, herein referred to as a risk aversion parameter. In our quantitative framework, we abstract from the construction of such a mapping, effectively assuming that the client communicates directly a single risk aversion parameter to the robo-advisor.\footnote{The majority of robo-advisors elicit risk preferences by means of online questionnaires (\cite{Beketov}). The client is presented with questions regarding, e.g., demographics, investment goals, education and financial literacy, and potential reactions to hypothetical gambles and market events. We refer to \cite{Charness} for an outline of the pros and cons of different methods used to assess risk preferences, and \cite{Cox} for a more comprehensive overview.} Note that between consecutive interaction times, the robo-advisor receives no input from the client.


The interaction schedule, denoted by  $(T_k)_{k\geq 0}$, is an increasing sequence of stopping times with respect to the filtration $(\calF_n)_{n\geq 0}$, defined in (\ref{calF}). That is, $T_0=0$, $T_k<T_{k+1}$, and 
$\{T_k\leq n\} \in \calF_{n}$, for any $n\geq 0$. Hence, interaction can be triggered by any combination of client-specific events, changes in the state of the economy, and market events such as a cascade of negative market returns. 

The rule used to determine the interaction schedule $(T_k)_{k\geq 0}$ is decided at the beginning of the investment process, and used throughout it. 
For future reference, we also define the process $(\tau_n)_{n\geq 0}$, where $$\tau_n := \sup\{T_k: T_k\leq n\},$$ is the last interaction time prior to and including time $n$. 


\subsection{Client's Risk Aversion Process}\label{SectionModelclient}

We first introduce the \emph{client's actual risk aversion process}, $(\gamma^C_n)_{n\geq 0}$, which is an $\bbr^+_0$-valued stochastic process adapted to the filtration $(\calF_n)_{n\geq 0}$ defined in (\ref{calF}). 
{This means that at} time $n$, the client's risk aversion may have shifted from its initial value due to changes in economic regimes, $Y_{(n)}$, and realized market returns $Z_{(n)}$, as well as because of idiosyncratic shocks to the client's risk aversion, $\epsilon_{(n)}$.

We {also} introduce the $(\calF_n)_{n\geq 0}$-adapted process $(\xi_n)_{n\geq 0}$, where $\xi_n\in\bbr_0^+$ is the risk aversion {parameter} \emph{communicated by the client} at the most recent interaction time, $\tau_n$. 
Observe that the process $(\xi_n)_{n\geq 0}$ depends on the interaction schedule $(T_k)_{k\geq 0}$ and, by construction, it remains constant between consecutive {interaction} times where no new information {comes} from the client, {i.e.,} $\xi_{n} = \xi_{\tau_n}$. 

The robo-advisor then constructs a \emph{model of the client's risk aversion process}, denoted by $(\gamma_n)_{n\geq 0}$, and uses it to solve the optimal investment problem. While it is generally desirable for the model $(\gamma_n)_{n\geq 0}$ to accurately track the client's actual risk aversion, $(\gamma_n^C)_{n\geq 0}$, the two {processes} may not coincide due to 
{the following reasons}. 
First, the client and the robo-advisor may not {interact at} all times, so the robo-advisor does not always have access to up-to-date information about the client's risk preferences. Specifically, while the robo-advisor observes both market returns and 
{changes in the economic regime}, {it cannot observe} {in real time} the idiosyncratic shocks to the client's risk aversion. 
Second, even 
if interaction were to occur at all times, information communicated by the client may not be representative of the client's true risk preferences due to her behavioral biases (see Section \ref{sectionModel} for further details). 


Formally, the risk aversion process $(\gamma_n)_{n\geq 0}$ is an $\bbr^+_0$-valued stochastic process adapted to the \emph{robo-advisor's filtration}, $(\calF_n^R)_{n\geq 0}$, 
which is generated by the random variables $(D_n)_{n\geq 0}$, defined as
\begin{align}\label{Dn}
\begin{split}
D_{n} &:= (Y_{(n)},Z_{(n)},\tau_{(n)},\xi_{(n)}) \in\mathcal{D}_n, \qquad
\mathcal{D}_n := \mathcal{Y}^{n+1} \times \bbr^n \times \bbn^{n+1} \times (\bbr_0^+)^{n+1}. 
\end{split}
\end{align}
It then follows that at time $n$, the risk aversion process may be written as
\begin{align}\label{gamma1}
\gamma_n:=\gamma_n(D_{n})\in\calF_n^R, 
\end{align}
for a measurable function $\gamma_n:\mathcal{D}_n\mapsto\bbr_0^+$.

\begin{rem}
	The random variable $D_n$ in (\ref{Dn}) can be decomposed as $D_n=(M_{n},I_{n})$, where $M_{n}:=(Y_{(n)},Z_{(n)})$, and $I_{n}:=(\tau_{(n)},\xi_{(n)})$. This shows that the robo-advisor's filtration $(\calF_n^R)_{n\geq 0}$ has two sources of information. 
	{The first component, $M_n$, reflects the fact that}
	the robo-advisor has the ability to process all available information about the market and the economy.
	{The second component, $I_n$, is the result of} 
	 its interaction with the client, and contains both the history of interaction times and the communicated risk aversion values. Under mild conditions, the robo-advisor's filtration grows with the frequency of interaction, and, {if interaction occurs} at all times, it becomes equal to the filtration $(\calF_n)_{n\geq 0}$.\footnote{For example, assume the interaction times to be given by $T_k=k\phi$, for some $\phi\geq 1$. If the sum of idiosyncratic shocks, $(\sum_{k=1}^{n}\epsilon_k)_{n\geq 0}$ is measurable with respect to the robo-advisor's filtration, then $(\calF_n^R)_{n\geq 0}$ increases to $(\calF_n)_{n\geq 0}$, as $\phi\downarrow 1$.} \hfill\qed	
\end{rem}

\subsection{Investment Criterion}\label{SectionInvestment}

The robo-advisor's objective is to optimally invest the client's wealth, 
taking into account the stochastic nature of the client's risk preferences. For this purpose, we develop 
a dynamic version of the standard \cite{Markowitz} mean-variance problem that adapts to the client's changing risk preferences. 
We proceed to introduce this criterion, which we refer to as an \emph{adaptive mean-variance criterion.}

Let $T\geq 1$ be a fixed investment horizon. For each $n\in\{0,1,\dots,T-1\}$, {$x\in\bbrZ$}, $d\in\mathcal{D}_n$, and a control law $\pi:=(\pi_n)_{n\geq 0}$, we consider the mean-variance functional 
\begin{align}\label{BjorkCriterion}
J_{n}(x,d;\pi) &:= \bbe_{n,x,d}[r_{n,T}^{\pi}] - \frac{\gamma_n(d)}{2} Var_{n,x,d}[r_{n,T}^{\pi}], 
\end{align}
where, at time $n$, the \emph{risk-return tradeoff} $\gamma_n$ is the robo-advisor's model of the client's risk aversion {at this time}, and $r_{n,T}^{\pi}$ is the simple return obtained by following the control law {$\pi$} until the terminal date $T$,
\begin{align}\label{return}
r_{n,T}^{\pi} := \frac{X^{\pi}_T-X_n}{X_n}.
\end{align}
The initial condition in (\ref{BjorkCriterion}) is given by $X_{n}=x$ and $D_n=d$, and both the expectation and the variance are computed with respect to the probability measure $\bbp_{n,x,d}(\cdot) := \bbp(\cdot|X_{n}=x,D_{n}=d)$. For future reference, we also {introduce} the probability measure $\bbp_{n,d}(\cdot) := \bbp(\cdot|D_{n}=d)$. 

Observe that the risk-return tradeoff $\gamma_n$ does not depend on the client's wealth. There are important reasons behind this assumption. 
First, it is consistent with the Markowitz mean-variance criterion, which is recovered as a special case of our model when there is a single investment period. 
Second, 
{as we will see in Section} \ref{SectionOptimalAllocation}, the optimal investment strategy turns out to be consistent with {that of} an investor exhibiting constant relative risk aversion (CRRA), i.e., {whose proportion of wealth invested in the risky asset} is independent of the wealth level.\footnote{Empirical evidence suggests that CRRA is a good description of microeconomic behavior (see, e.g., extensive panel data studies carried out in \cite{Nagel1,Chiappori}, and \cite{Sahm}). However, the evidence is not universal. For instance, 
	\cite{Calvet} and \cite{GuisoPaiella} reject CRRA in favor of decreasing relative risk aversion.}

The objective functional (\ref{BjorkCriterion}), together with the risk aversion process (\ref{gamma1}), define a family of \emph{sequentially adaptive} {optimization} problems, in the sense that at each time a new problem arises, with properties that depend on realized market returns, economic state changes, and client-communicated information, but with the same initially {specified} terminal date. 
Moreover, observe that it is the robo-advisor that solves the optimization problem, and the initial condition of each optimization problem fixes the value of the stochastic process that generates the robo-advisor's filtration, as well as the value of the client's wealth process.

{The} control law $\pi$ in (\ref{BjorkCriterion}), also referred to as a strategy or allocation, is such that for each $n$, the control $\pi_n$ is a measurable real-valued function of the state variables $X_n$ and $D_n$.  
Additionally, admissible control laws are assumed to be self-financing 
and to satisfy the square-integrability condition
$\bbe\big[\sum_{n=0}^{T-1}\pi_n^2\big] < \infty$.

\begin{rem}
	We assume the investment horizon $T$ to be deterministic and fixed. A stochastic {horizon} can be captured through {distributional assumptions on} the  risk aversion process $(\gamma_n)_{n\geq 0}$. 
	{For example, the possibility of {client death} can be modeled by introducing} a time-dependent probability $p_n$ of the risk aversion parameter $\gamma_{n+1}$ becoming ``infinite''. The probability $p_n$ would be increasing in $n$, and infinity would be an absorbing state reached only in the event {that the client dies.} In the death-state, the client's portfolio is liquidated {and the client's entire wealth} is  allocated to the risk-free asset. For the optimization problem described in this section, a non-zero value of $p_n$ would have the effect of tilting the risky asset allocation upward, {because no more investing is possible after death.} {In the same vein as when accounting for client death, we may impose conditions on the dynamics of the risk aversion process that halt investment if certain market events or economic conditions occur.} 
	\hfill\qed
\end{rem}


\begin{rem} 
	There is a non-zero probability that the client's wealth becomes negative. This can be either due to short-selling or to an extreme negative return of the risky asset (see Remark \ref{S0neg}). Although the objective functional (\ref{BjorkCriterion}) and the optimal strategy presented in Section \ref{SectionOptimalAllocation} are well-defined for negative wealth levels, they are not economically meaningful and, in practice, the robo-advisor would liquidate the client's portfolio as soon as the wealth becomes negative. 
	
	In Appendix \ref{secConstraints}, we {modify the optimal strategy in a way that forces liquidation of the entire risky position if the wealth is negative.} We show that the change in expectation and variance of the optimal portfolio's return (given in (\ref{return})), resulting from imposing this liquidation condition, can be bounded by a term that is extremely small for plausible values of the model parameters. {It then follows that forcing liquidation has a small impact on the value of the objective functional} (\ref{BjorkCriterion}).
		
	The liquidation condition is imposed {ad hoc} on the optimal strategy. Hence, it is myopic in the sense that {the modified strategy} coincides with the optimal strategy until the {wealth achieves a negative value}. In Appendix \ref{secConstraints}, we also discuss how the optimal investment problem can be adjusted to handle portfolio constraints which are typically imposed by robo-advising firms, such as no borrowing and no short-selling. In that case, the constraints also affect the optimal strategy at times before they become binding, because of the dynamic nature of the optimization problem. 	 
	\hfill\qed  
\end{rem}

Through the wealth dynamics (\ref{dX}), the mean-variance functional $J_n$ depends on the control law $\pi$ restricted to the time points $\{n,n+1,\dots,T-1\}$ and the robo-advisor chooses the control $\pi_{n}$ \emph{given} future control decisions $\pi_{n+1:T}:=\{\pi_{n+1},\pi_{n+2},\dots,\pi_{T-1}\}$. Therefore, any candidate optimal control law $\pi^*$ is such that for each $n\in\{0,1,\dots,T-1\}$, {$x\in\bbrZ$}, and $d\in\mathcal{D}_n$,
\begin{align}\label{BjorkCriterion2}
&\sup_{\pi\in A_{n+1}^*}J_{n}(x,d;\pi)=J_{n}(x,d;\pi^*),
\end{align}
where $A_{n+1}^*:=\{\pi:\pi_{n+1:T}=\pi_{n+1:T}^*\}$ is the set of control laws that coincide with $\pi^*$ after time $n$. 
If a control law $\pi^*$ satisfying (\ref{BjorkCriterion2}) exists, we define the corresponding value function at time $n$ as 
\begin{align}\label{V}
V_{n}(x,d) := J_{n}(x,d;\pi^*).
\end{align} 

{It is worth noticing that when seeking an optimal control $\pi^*_n$ at time $n$, the robo-advisor takes into account the dynamics of $\{\gamma_n,\gamma_{n+1},\dots,\gamma_{T-1}\}$, i.e., the dynamics of the risk-return coefficient \emph{throughout} the investment horizon, with only the \emph{current} value $\gamma_n$ being known. At the subsequent time $n+1$, the dynamics of the remaining values $\{\gamma_{n+1},\dots,\gamma_{T-1}\}$ are updated by the robo-advisor, and $\gamma_{n+1}$ becomes known. 
If $n+1$ is a non-interaction time, this update is based only on information about market returns and economic states, while if $n+1$ is a time of interaction it also accounts for information received from the client.}

\section{Optimal Investment Strategy}\label{SectionOptimalAllocation}

We analyze the optimization problem defined by the sequence of objective functionals $(J_n)_{0\leq n<T}$ in  (\ref{BjorkCriterion}) and the optimality criterion in (\ref{BjorkCriterion2}). It is well known that, even if the risk-return tradeoff $(\gamma_n)_{n\geq 0}$ is constant {through} time, the family of optimization problems defined by  (\ref{BjorkCriterion}) is time-inconsistent in the sense that the Bellman optimality principle does not  hold. This means that if, at time $n$, the control law $\pi^*$ maximizes the objective functional $J_{n}$, then, at time $n+1$, the restriction of $\pi^*$ to the time points $\{n+1,n+2,\dots,T-1\}$ may not maximize $J_{n+1}$.
We refer to \cite{Bjork} and references therein for a general framework of time-inconsistent stochastic control in discrete time. 

As standard in this literature, 
we view the optimization problem in (\ref{BjorkCriterion2}) as a multi-player game, where the player at each time $n\in\{0,1,\dots,T-1\}$ is thought of as a future self of the client. Player $n$ then wishes to maximize the objective functional $J_{n}$, but decides only the strategy $\pi_n$ at time $n$, while $\pi_{n+1},\dots,\pi_{T-1}$ are determined by her future selves. The resulting optimal control strategy, $\pi^*$, is the subgame perfect equilibrium of this game, and can be computed using backward induction. At time $n=T-1$, the equilibrium control $\pi^*_{T-1}$ is obtained by maximizing $J_{T-1}$ over $\pi_{T-1}$, which is a standard single-period optimization problem. For $n<T-1$, the equilibrium control $\pi^*_{n}$ is then obtained by letting player $n$ choose $\pi_{n}$ to maximize $J_{n}$, given that player $n'$ will use $\pi^*_{n'}$, for $n'=n+1,n+2,\dots,T-1$.

In Appendix \ref{AppendixA}, we derive an extended Hamilton-Jacobi-Bellman (HJB) system of equations satisfied by the value function of the optimization problem. The following result presents the solution to this system, which is the optimal investment strategy for an interaction schedule $(T_k)_{k\geq 0}$ and a risk aversion process $(\gamma_n)_{n\geq 0}$ of the general form, respectively introduced in Sections \ref{SectionModelInteraction} and \ref{SectionModelclient}.

	
\begin{prop}\label{prop1}
	The dynamic optimization problem (\ref{BjorkCriterion2}) admits a solution $\pi^*$ of the form
	\begin{align*}
	\pi_n^*(x,d) = \tilde\pi_n^*(d)x, \quad 0\leq n<T, 
	\end{align*}
	for {$x\in\bbrZ$} and $d\in\mathcal{D}_n$. 
	The optimal proportion of wealth allocated to the risky asset is given by 
	\begin{align}\label{pi_hat}
	\tilde\pi^*_n(d) &= \frac{1}{\gamma_n}\frac{\bbe_{n,d}[\widetilde Z_{n+1}\tilde r_{n+1,T}^{\pi^*}]}{Var_{n,d}[\widetilde Z_{n+1}\tilde r_{n+1,T}^{\pi^*}]}
	- R_{n+1}\frac{Cov_{n,d}[\widetilde Z_{n+1}\tilde r_{n+1,T}^{\pi^*},\tilde r_{n+1,T}^{\pi^*}]}{Var_{n,d}[\widetilde Z_{n+1}\tilde r_{n+1,T}^{\pi^*}]}, 
	\end{align}
	where we recall that $R_{n+1}$ and $\widetilde Z_{n+1}$ have been defined in Section \ref{SectionModelRobo}, and \begin{align}\label{rnT}
	\tilde r_{n+1,T}^{\pi^*}:=1+ r_{n+1,T}^{\pi^*} = \frac{X_T^{\pi^*}}{X_{n+1}},
	\end{align}  
	is the terminal value of one dollar invested in the optimal strategy $\pi^*$ at time $n+1$. 
\end{prop}

The above proposition characterizes the structure of the optimal strategy. It shows that the optimal allocation, i.e., the relative fraction of wealth allocated to each asset, is independent of the wealth level. 
In Appendix \ref{Appendix0}, we use backward induction to {develop an explicit representation} of the optimal strategy, and discuss {the} computational complexity {of this procedure.} 
As a byproduct, we obtain formulas for the expected value and variance of 
the optimal portfolio's return, which in turn can be used to compute the value function (\ref{V}). 

It is evident from (\ref{pi_hat}) that the optimal strategy can be decomposed into two terms. The first term resembles the standard single-period Markowitz strategy, $\tilde\mu_{n+1}/(\gamma_n\sigma_{n+1}^2)$. However, rather than depending only on statistics of the market return between $n$ and $n+1$, it also accounts for the return $\tilde r_{n+1,T}^{\pi^*}$ achieved by the optimal portfolio between times $n+1$ and $T$. The second term in the decomposition is the \emph{intertemporal hedging demand}, which can be rewritten as
\begin{align*}
Cov_{n,d}\big[\widetilde Z_{n+1}\tilde r_{n+1,T}^{\pi^*},\tilde r_{n+1,T}^{\pi^*}\big]
&=Cov_{n,d}\big[(\widetilde Z_{n+1}-\tilde\mu_{n+1})\tilde r_{n+1,T}^{\pi^*},\tilde r_{n+1,T}^{\pi^*}\big]
+\tilde \mu_{n+1}Var_{n,d}(\tilde r_{n+1,T}^{\pi^*}),
\end{align*}
where the first term captures the contribution of dynamic risk aversion. To analyze this term, it is convenient to first rewrite it as
\begin{align*}
Cov_{n,d}\big[(\widetilde Z_{n+1}-\tilde\mu_{n+1})\tilde r_{n+1,T}^{\pi^*},\tilde r_{n+1,T}^{\pi^*}\big]
&=Cov_{n,d}\big[(\widetilde Z_{n+1}-\tilde\mu_{n+1})\bbe_{n+1,D_{n+1}}[\tilde r_{n+1,T}^{\pi^*}],\bbe_{n+1,D_{n+1}}[\tilde r_{n+1,T}^{\pi^*}]\big],
\end{align*}
which shows that its sign depends predominantly on the covariance between $\widetilde Z_{n+1}-\tilde\mu_{n+1}$, i.e., the \emph{current} excess market return, and $\bbe_{n+1,D_{n+1}}[\tilde r_{n+1,T}^{\pi^*}]$, i.e., the \emph{future} expected portfolio return.
This covariance is generally positive, resulting in a negative hedging demand. 
To see this, observe that we generally expect an above average market return, $\widetilde Z_{n+1}-\tilde\mu_{n+1}>0$, to push down the future risk aversion values $(\gamma_k)_{n+1\leq k<T}$; this, in turn, implies a positive relation between $\widetilde Z_{n+1}-\tilde\mu_{n+1}$ and future risky asset allocations $(\tilde\pi_{k}^*)_{n+1\leq k<T}$. As a result, the expectation $\bbe_{n+1,D_{n+1}}[\tilde r_{n+1,T}^{\pi^*}]$, 
achieved through the allocations $(\tilde\pi_{k}^*)_{n+1\leq k<T}$, is larger following a positive value of $\widetilde Z_{n+1}-\tilde\mu_{n+1}$. 
Altogether, the current market return and the future expected portfolio return thus move in the same direction, amplifying the effect of the current market return on the variance of the terminal wealth and leading to a negative hedging demand.
	

This is akin to the intertemporal hedging term in \cite{Basak}, which arises because of covariation between market returns and the state variable.
In their setting, the hedging demand would vanish if market returns and changes in the state variable were uncorrelated, in which case trading in the risky asset cannot hedge fluctuations in the state variable. In our setting, the hedging demand is driven by the dynamic risk aversion process that links current market return with future portfolio returns. This hedging demand appears even though the market returns in our model are conditionally independent of changes in the state variable.
\footnote{Conditionally on the value of $Y_n$, the market return $\widetilde Z_{n+1}$ is independent of $Y_{n+1}-Y_n$.}


Next, we consider the special case where intertemporal hedging due to dynamic risk aversion vanishes, i.e., when future risk aversion values are conditionally independent of the current market return. 
In this case, the optimal allocation strategy admits a more explicit expression, given only in terms of the first two moments of the return variables $\widetilde Z_{n+1}$ and $\tilde r_{n+1,T}^{\pi^*}$. 



\begin{cor}\label{cor1}
	Given $D_n=d$, assume the risk aversion values $(\gamma_{n+1}, \gamma_{n+2},\dots, \gamma_{T-1})$ to be independent of the 
	{market} return $\widetilde Z_{n+1}$. The optimal proportion of wealth allocated to the risky asset at time $n$ is given by
	\begin{align*}
	\tilde\pi^*_n(d) 
	&= \frac{1}{\gamma_n}\frac{\tilde\mu_{n+1}\bbe_{n,d}[\tilde r_{n+1,T}^{\pi^*}]}{\sigma_{n+1}^2\bbe_{n,d}[(\tilde r_{n+1,T}^{\pi^*})^2]+\tilde\mu_{n+1}^2Var_{n,d}[\tilde r_{n+1,T}^{\pi^*}]}
	- 
	R_{n+1}\frac{\tilde\mu_{n+1}Var_{n,d}[\tilde r_{n+1,T}^{\pi^*}]}{\sigma_{n+1}^2\bbe_{n,d}[(\tilde r_{n+1,T}^{\pi^*})^2]+\tilde\mu_{n+1}^2Var_{n,d}[\tilde r_{n+1,T}^{\pi^*}]}, 
	\end{align*}
	where $\tilde\mu_{n+1}=\bbe_{n,d}[\widetilde Z_{n+1}]$ and $\sigma_{n+1}^2=Var_{n,d}[\widetilde Z_{n+1}]$. 
\end{cor}
The optimal allocation given in the corollary can alternatively be written as\footnote{An extension of formula (\ref{u_approx}) can be derived for the general model {considered} in Proposition 4.1 (see Appendix \ref{AppendixA}).}
\begin{align}\label{u_approx}
\begin{split}
\tilde\pi^*_n(d) 
&=\frac{\tilde\mu_{n+1}}{\gamma_n\sigma_{n+1}^2}\frac{\bbe_{n,d}[\tilde r_{n+1,T}^{\pi^*}]-R_{n+1}\gamma_n\big(\bbe_{n,d}[(\tilde r_{n+1,T}^{\pi^*})^2]-(\bbe_{n,d}[\tilde r_{n+1,T}^{\pi^*}])^2\big)}{\bbe_{n,d}[(\tilde r_{n+1,T}^{\pi^*})^2] + \Big(\frac{\tilde\mu_{n+1}}{\sigma_{n+1}}\Big)^2\big(\bbe_{n,d}[(\tilde r_{n+1,T}^{\pi^*})^2]- (\bbe_{n,d}[\tilde r_{n+1,T}^{\pi^*}])^2\big)},
\end{split}
\end{align}
which allows to pin down the relation with the optimal investment strategy in the classical mean-variance setup. Specifically, it is proportional to the single-period Markowitz strategy, $\tilde\mu_{n+1}/(\gamma_n\sigma_{n+1}^2)$, i.e., the optimal allocation of a myopic investor whose objective functional spans a single period. 
This fraction depends on the \emph{current} economic conditions, and the client's \emph{current} risk aversion. 
The proportionality factor in (\ref{u_approx}) depends on the \emph{future} return of the investment strategy, $\tilde r_{n+1,T}^{\pi^*}$, 
which {in turn} depends on both future economic conditions and the client's future risk aversion dynamics. 
Consistently with intuition, this factor is increasing in the first moment of $\tilde r_{n+1,T}^{\pi^*}$ and decreasing in {its} second moment. 
{In} the final time period, the proportionality factor is equal to one and the optimal allocation {is given explicitly} by the Markowitz strategy in this time period.

\section{Performance of the Robo-Advising Framework}\label{SectionRegret}



We analyze the investment performance of the proposed robo-advising framework. Our analysis will be based on a specific model that fits into the general framework, and which we introduce in Section~\ref{sectionModel}. In Section \ref{SectionRegretDef1}, we consider the interplay between interaction and portfolio personalization, {and show the existence of a tradeoff between the \emph{rate of information acquisition} from the client and the \emph{effect of the client's behavioral biases}.} 
In Section \ref{SectionY}, we explore how the optimal investment strategy is affected by transitions between economic states and the associated changes in the client's risk aversion. We also discuss whether the robo-advisor should cater to client wishes or go against them to improve the investment performance. 

\subsection{A Robo-Advising {Model}}\label{sectionModel}


{In the following paragraphs, we specify each component of the framework introduced in Sections} \ref{SectionModelRobo}-\ref{SectionInvestment}. 

\paragraph{Market Dynamics.} The risky asset has conditionally Gaussian returns. That is, given the economic state $Y_n=y$ at time $n$,
the return $Z_{n+1}$ has a Gaussian distribution with mean $\mu(y)$ and variance $\sigma^2(y)$. 

\paragraph{State Process Dynamics.} We assume a two-state economy $(Y_n)_{n\geq 0}$. This choice is supported by the methodology of the National Bureau of Economic Research (NBER), which splits business cycles into periods of economic expansions and contractions.\footnote{The work of \cite{Chauvet} {shows that a Markov switching model successfully identifies NBER's business cycle turning points.}} We use $y=1$ {to denote the state of} economic growth and $y=2$ {to denote the state of recession}. We use $P$ to denote the transition probability matrix, so that $P_{ij}$ is the probability that the economy transits from state $i$ to state $j$ in a single time step. 

\paragraph{Interaction Schedule.} The sequence of interaction times is deterministic and equally spaced. The interaction schedule $(T_k)_{k\geq 0}$ is then characterized by a fixed integer parameter $\phi\geq 1$, with $T_k=k\phi$. The two extreme cases are $\phi=1$, which corresponds to communication of risk preferences at all times, and $\phi\geq T$, which corresponds to risk preferences {communicated only} at the beginning of the investment process. 

\paragraph{Client's Risk Aversion.}

The client's risk aversion process, $(\gamma_n^C)_{n\geq 0}$, is of the form
\begin{align}\label{gammaCex}
\gamma_n^C = e^{\eta_n}\gamma_n^{id}\gamma_n^{Y}.
\end{align} 
{The first component, $(e^{\eta_n})_{n\geq 0}$, is deterministic and time-varying.} 
For example, the specification $e^{\eta_n} = e^{-\alpha(T-n)}$ captures the empirical observation that risk aversion increases with age, with the parameter $\alpha\geq 0$ specifies the rate of increase.

{The second component, $(\gamma_n^{id})_{n\geq 0}$, depends on the client's personal circumstances and has stochastic dynamics,}
{$\gamma_n^{id} = \gamma_{n-1}^{id}e^{\epsilon_n}$, where $(\epsilon_n)_{n\geq 1}$ is an} i.i.d.\ sequence {of random variables. Specifically,  $\epsilon_n=\sigma_{\epsilon}W_n-\frac{\sigma_{\epsilon}^2}{2}$ with probability $p_{\epsilon}\in[0,1]$, where $\sigma_{\epsilon}>0$ and $(W_n)_{n\geq 1}$ is an} i.i.d.\ {sequence of standard Gaussian random variables, and $\epsilon_n=0$ with probability $1-p_{\epsilon}$.}
This component captures idiosyncratic shocks to the client's risk aversion, which are unrelated to economic state transitions and market dynamics.   
The multiplicative innovation terms $(e^{\epsilon_n})_{n\geq 1}$ are independent {and with unit mean}, making $(\gamma_n^{id})_{n\geq 0}$ a martingale, devoid of a predictable component. 


The third component, $(\gamma_n^Y)_{n\geq 0}$, is given by {$\gamma_n^Y=\bar\gamma_n(Y_n)$,}
where $\bar\gamma_n:\mathcal{Y}\mapsto\bbr_0^+$ is a possibly time-varying function of the current economic state. {This component captures the empirical fact that risk aversion varies with the business cycle.} 


\paragraph{Client's Behavioral Bias.} At time $n$, the risk aversion parameter communicated by the client 
at the most recent interaction time, $\tau_n$, is of the form 
\begin{align}\label{gammaCZ}
\xi_n = \gamma_{\tau_n}^{C}\gamma_{\tau_n}^{Z} := \gamma_{\tau_n}^{C}e^{-\beta\big(\frac{1}{\phi}\sum_{k=\tau_n-\phi}^{\tau_n-1}(Z_{k+1}-\mu_{k+1})\big)},
\end{align}
where $\gamma_{\tau_n}^{Z}$ is a factor that inflates or deflates the client's actual risk aversion, $\gamma_{\tau_n}^{C}$, based on recent market returns. 
Namely, the sum in the exponent is the cumulative excess return of the risky asset over its expected value, since the previous interaction time, $\tau_n-\phi$. 
Because previous market returns have no predictive value for future market returns, given the state of the economy, the factor $\gamma_{\tau_n}^Z$ is representative of common behavioral biases, and the coefficient $\beta\geq 0$ {determines} the magnitude of this effect. 

We emphasize two important properties of the component $\gamma_{\tau_n}^Z$. First, the convexity of the exponential function allows us to capture loss aversion: the upward risk aversion bias following a market underperformance is greater than the downward bias when the market exceeds expectations by the same amount (\cite{Kahneman1}).
Second, in a model with a single economic state $y$, we have $\bbe[\gamma_{\tau_n}^Z] = e^{\frac{\beta\sigma^2(y)}{2\phi}}$,  which shows that the average effect of the behavioral bias decreases as the time between interaction increases,\footnote{In a setup with multiple economic states we have, conditionally on $Y_{\tau_n-\phi}=y\in\mathcal{Y}$, that $\bbe[\gamma_{\tau_n}^Z] = e^{\frac{\beta\sigma^2(y)}{2\phi}}+O(1-P_{y,y}^{\phi})$.} consistent with the notion of myopic loss aversion, introduced in \cite{Benartzi}.

	The {bias} $\gamma_{\tau_n}^Z$ may be {interpreted} as the client overriding the robo-advisor's decisions. That is, at an interaction time $n$, the robo-advisor proposes a portfolio 
	{tailored to} the client's characteristics, $\gamma_{n}^C$, but the client makes changes to {the} proposed allocation based on recent market returns. This results in a portfolio allocation {consistent with a risk aversion coefficient} $\gamma_{n}^C\gamma_{n}^Z$. The effect of this {override}, which is {driven by} the client's behavioral biases, prevails until the subsequent time of interaction, $n+\phi$, when the client again adjusts the portfolio allocation proposed by the robo-advisor. This time, however, the adjustment is based on market returns {realized} in the time interval $[n,n+\phi]$.

{\paragraph{Robo-Advisor's Model of Client's Risk Aversion.} The risk aversion process $(\gamma_n)_{n\geq 0}$ is of the form
	\begin{align}\label{gammaEx}
	\gamma_n  = e^{\eta_n-\eta_{\tau_n}}\xi_{n}\frac{\gamma_n^Y}{\gamma_{\tau_n}^Y}. 
	\end{align}
	The component $\xi_n$ is the risk aversion parameter communicated by the client at the most recent interaction time, $\tau_n$, and, in the robo-advisor's model, it stays constant until the subsequent time of interaction, $\tau_n+\phi$. On the other hand, the model is adjusted for the passage of time through the factor $e^{\eta_n-\eta_{\tau_n}}$, and for changes in 
	economic states through the ratio $\gamma_n^Y/\gamma_{\tau_n}^Y$. The above equation can be written in an alternative form as
	\begin{align*} 
	\gamma_n = \gamma_n(d) = \gamma_{\tau_n}^Z \bbe_{n,d}[\gamma_n^C], 
	\end{align*}
	where we recall from (\ref{Dn}) that $D_n=d$ is the information set of the robo-advisor at time $n$. 
	This shows that the robo-advisor's view of the client risk aversion is 
	equal to the expectation of client's actual risk aversion 
	times bias factor from the previous interaction time.
	
	{From the above expression it can be observed that interaction impacts the robo-advisor's inferred value of the client's risk aversion in two different ways. First, between interaction times, the expectation $\bbe_{n,d}[\gamma_n^C]$ is not updated in response to changes in the idiosyncratic component of the client's risk aversion, because such changes cannot be inferred from observables like market returns or economic indicators.}
	{Second, a higher interaction frequency comes at the expense of increased behavioral bias, tilting the robo-advisor's model further away from the client's actual risk aversion.}

	\paragraph{Calibration of Risk {Aversion Process}.}\label{sec:modelcalibration}
	The multiplicative model introduced in this section provides a parsimonious description of the client's risk aversion. Under this specification, the risk aversion is guaranteed to be non-negative, and its parameters can be estimated from historical portfolio allocations. We next sketch a calibration procedure for models of this form.

	Consider a sequence of historical portfolio allocations for a set of self-directed investors. Such data is available to financial firms offering discount brokerage services, in particular the growing number of low-cost online brokerages.\footnote{See, among others, the seminal work of \cite{Odean} where such a data set is analyzed and \cite{Glaser} {for a more recent study using online brokerage data.}}
	Assume that the portfolio allocations are consistent with a myopic mean-variance criterion, for which there exists a one-to-one relationship between allocations and risk aversion levels. Then, the allocations can equivalently be viewed as historical sequences of implied risk aversion coefficients.
	
	For a given client, the risk aversion model can be calibrated using the subset of investors that have demographic characteristics similar to the client.\footnote{{In practice, robo-advising firms manage a finite number of portfolios and group clients with similar profiles into risk categories. For instance, Wealthfront constructs a composite Risk Score ranging from 0.5 to 10, in increments of 0.5.}}	
	Specifically, the three components of the risk aversion process (\ref{gammaCex}) and the behavioral bias in (\ref{gammaCZ}) can be estimated using the following sequential procedure. 
	First, the temporal component can be inferred from the initial portfolio composition chosen by investors at different life stages. For a personalized service, this component can also be used to incorporate a client-specific time-dependence of risk aversion. Adjusting for the effect of time, the average investment levels during each economic regime (e.g., periods of expansions and contractions - see Section \ref{SectionY}) then provide information about the process component which depends on the economic state. Next, the extent of behavioral bias can be estimated from 
	the relationship between market returns and subsequent changes in the portfolio's market exposure.\footnote{{The literature on individual investor performance focuses to a large extent on biases and investor's mistakes. For instance,}  \cite{Statman} {study the role of investor overconfidence by examining the lead-lag relationship between market returns and trading volume.}} In addition, {for a given client,} data on financial literacy, investment experience, and cognitive abilities, can be used to adjust the estimated behavioral bias (see, e.g., studies by \cite{Oechssler} and \cite{Seasholes}). 
	Finally, {after adjusting} for {the} passage of time, economic regimes, and bias, the size and frequency of allocation changes made by investors provide information about the idiosyncratic component of the risk aversion process.

\paragraph{Sensitivity of Optimal Allocation to Idiosyncratic Shock and Behavioral Bias.}
	
	In Figure \ref{fig_jumps2_beta}, {we display how the optimal allocation in Proposition} \ref{prop1} depends on the variability of the idiosyncratic component of the client's risk aversion and the client's behavioral bias. 
	The left panel shows that if the client's idiosyncratic risk preferences {fluctuate more due to a larger} value of the parameter $p_{\epsilon}$, then the optimal allocation deviates more from that of the benchmark case with no idiosyncratic shocks. The right panel shows 
	that the client's behavioral bias following a negative market return ($\gamma_n^Z>1$) 
	has a stronger influence on the optimal allocation than the bias following a positive market return of the same magnitude ($\gamma_n^Z<1$). This numerical finding is consistent with the notion of loss aversion, as discussed above.

\begin{figure}
	\centering
	\includegraphics[width=\textwidth]{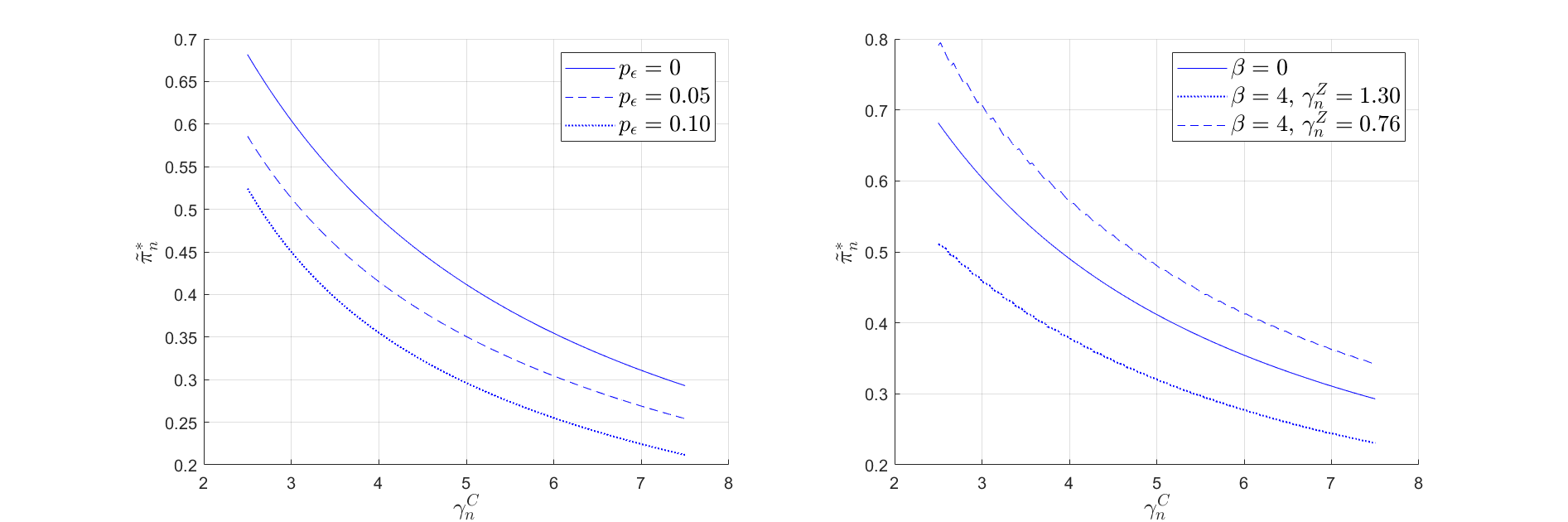}
	\caption{\small  
		The graphs plot the optimal allocation $\tilde\pi^*_n$ as a function of the client's risk aversion $\gamma_n^C$, for the model in Section \ref{sectionModel}, where we set $Y_0=1$ and $P$ equal to the two-dimensional identity matrix. In both graphs, the solid line refers to the benchmark case of constant risk aversion and no behavioral bias, i.e., $\eta_n=p_{\epsilon}=\beta=0$. 
		\newline Left panel: Impact of idiosyncratic risk aversion shocks, obtained by letting $p_{\epsilon}\in\{0.05,0.10\}$ and $\sigma_{\epsilon}=0.64$. 
		Right panel: Impact of behavioral bias, obtained by choosing $\beta=4$ and $\gamma_n^Z\in\{0.76,1.30\}$.
		These values of $\beta$ and $\gamma_n^Z$ correspond to a cumulative market return between two interaction times which is $1.3$ standard deviations above/below its expected value. The annualized market parameters in state $Y_0$ are set to $r(Y_0)=0$, $\mu(Y_0)=0.10$, $\sigma(Y_0)=0.20$, and one time step in the model corresponds to one month. The investment horizon is $T=36$ months, and the time between interaction is $\phi=3$ months. We show the allocations prevailing one year after the start of the investment process, i.e., at time $n=12$.}
\label{fig_jumps2_beta}
\end{figure}

\subsection{Interaction and Optimal Personalization}\label{SectionRegretDef1}



In our framework, the client and robo-advisor do not interact at all times and, thus, {the} robo-advisor is not always immediately aware of changes 
{in} the client's risk profile. While a higher interaction frequency is the only way to reduce this information asymmetry, 
it also increases the {amount of} behavioral bias in the communicated information. The goal of this section is to analyze this tradeoff quantitatively.

{We start by defining} a measure of portfolio personalization. This measure {depends on the relation} between the client's risk aversion process, $(\gamma_n^C)_{n\geq 0}$, given by (\ref{gammaCex}), and the robo-advisor's {model} of the client's risk aversion, $(\gamma_n)_{n\geq 0}$, given by (\ref{gammaEx}). 
Recall from Section \ref{SectionOptimalAllocation} that at time $n=T-1$, the optimal risky asset allocation is proportional to the risk tolerance parameter $1/\gamma_n$, and it is approximately proportional to $1/\gamma_n$ for $n<T-1$ (see (\ref{u_approx})).
We therefore choose to define personalization in terms of the expected relative difference between the risk tolerance parameters $1/\gamma_n$ and $1/\gamma_n^C$, averaged over the investment period. Specifically, for $\phi\geq 1$ and $\beta\geq 0$, we {introduce the \emph{personalization measure}}
\begin{align}\label{R}
\mathcal{R}(\phi,\beta)
&:=\bbe\left[\frac{1}{T}\sum_{n=0}^{T-1}\Big|\frac{\frac{1}{\gamma_n}-\frac{1}{\gamma_n^C}}{\frac{1}{\gamma_n^C}}\Big|\right]{.}
\end{align}
A lower value of $\mathcal{R}(\phi,\beta)$ means a higher level of portfolio personalization. {Full personalization is achieved if and only if $\phi=1$ and $\beta=0$, which leads to $\mathcal{R}(1,0)=0$. This is because the processes $(\gamma_n)_{n\geq 0}$ and $(\gamma_n^C)_{n\geq 0}$ coincide only in this case, when there is no behavioral bias and the client and the robo-advisor interact at all times.}
Although not explicitly highlighted in the notation, {we note that} the value of $\mathcal{R}(\phi,\beta)$ also depends on the parameters $p_{\epsilon}\in[0,1]$ and $\sigma_{\epsilon}>0$, which govern the distribution of idiosyncratic risk aversion shocks. 

For a given value of $\beta$, our objective is to study the dependence of the measure $\mathcal{R}$ on $\phi$. For this, we consider an approximation $\widetilde{\mathcal{R}}$ that is analytically more tractable {than $\mathcal{R}$}. In the following proposition, we characterize this approximation and show that it is minimized by a unique value of $\phi$. In Figure \ref{gamma_diff_approx}, we show graphically that $\widetilde{\mathcal{R}}$ and $\mathcal{R}$ are close, for different interaction frequencies and levels of behavioral bias.

\begin{prop}\label{propRegret}
	Set $\sigma_0:=\sigma(Y_0)$. The following statements hold:
	\begin{itemize} 
		\item[(i)]  
		The measure ${\mathcal{R}}(\phi,\beta)$ satisfies {the relation} 
		\begin{align}\label{genT}
		\frac{T_{\phi}}{T}\widetilde{\mathcal{R}}(\phi,\beta) + \mathcal{E}
		\leq \mathcal{R}(\phi,\beta) 
		\leq \frac{T^{\phi}}{T}\widetilde{\mathcal{R}}(\phi,\beta) + \mathcal{E},
		\end{align}
		where $T_{\phi}$ (resp.\ $T^{\phi}$) {is the largest (resp.\ smallest) multiple of $\phi$ that is less than (resp.\ greater than) or equal to $T$, the error term $\mathcal{E}$ is of order $O(p_{\epsilon}^2\sigma_{\epsilon}^2 + \beta^2) + O(1-P_{y_0,y_0}^{T-1})$, and}
		\begin{align*}
		\widetilde{\mathcal{R}}(\phi,\beta)
		&:=\sqrt{\frac{2}{\pi}}\Big(\frac{\beta\sigma_0}{\sqrt{\phi}}\Big(1  - \frac{\phi-1}{2}p_{\epsilon}\Big)
		+ \sqrt{\frac{\beta^2\sigma_0^2}{\phi}+\sigma_{\epsilon}^2}\frac{\phi-1}{2}p_{\epsilon}\Big).
		\end{align*}
		\item[(ii)]
		Let $\beta$ and $p_{\epsilon}$ be such that $\beta+p_{\epsilon}>0$. Then, there exists a unique value of $\phi\geq 1$ that minimizes $\widetilde{\mathcal{R}}(\beta,\phi)$.\footnote{If $\beta=p_{\epsilon}=0$, the values $\mathcal{R}(\phi,\beta)$ and $\widetilde{\mathcal{R}}(\phi,\beta)$ are independent of $\phi$.} Specifically,
		\begin{align*}
		\argmin_{\phi\geq 1} \widetilde{\mathcal{R}}(\phi,\beta) = \left\{\begin{array}{ll} 
		1, &\quad \beta=0, \\
		\phi_0\in[1,\infty),  &\quad \beta>0, \; p_{\epsilon}>0, \\
		\infty, &\quad p_{\epsilon} = 0.
		\end{array} \right.
		\end{align*}
		Furthermore, $\phi_0$ satisfies the monotonicity properties
		\begin{align}\label{deltas}
		\frac{\partial\phi_0}{\partial\beta}>0, \qquad \frac{\partial\phi_0}{\partial\sigma_0}>0, \qquad \frac{\partial\phi_0}{\partial p_{\epsilon}}<0, \qquad \frac{\partial\phi_0}{\partial\sigma_{\epsilon}}<0.
		\end{align} 
	\end{itemize}
\end{prop}

The error term of order $O(1-P_{y_0,y_0}^{T-1})$ in the approximation (\ref{genT}) results from fixing the economic state throughout the investment period. 
Typically, transitions between economic states are infrequent (see (\ref{Lambda}) in Section \ref{SectionY}) and {the error term vanishes if the model consists {of a single} economic state}. {From the signs of the derivatives given in} (\ref{deltas}), it can be seen that if the economy transits to a state with a higher return volatility, i.e., if $\sigma_0$ increases, the optimal time between consecutive interactions also increases. This is because the client's behavioral bias {is} based on market return fluctuations, which are magnified {if the return volatility is higher}. 



\begin{figure}[!ht]
	\centering
	\includegraphics[width=0.8\textwidth]{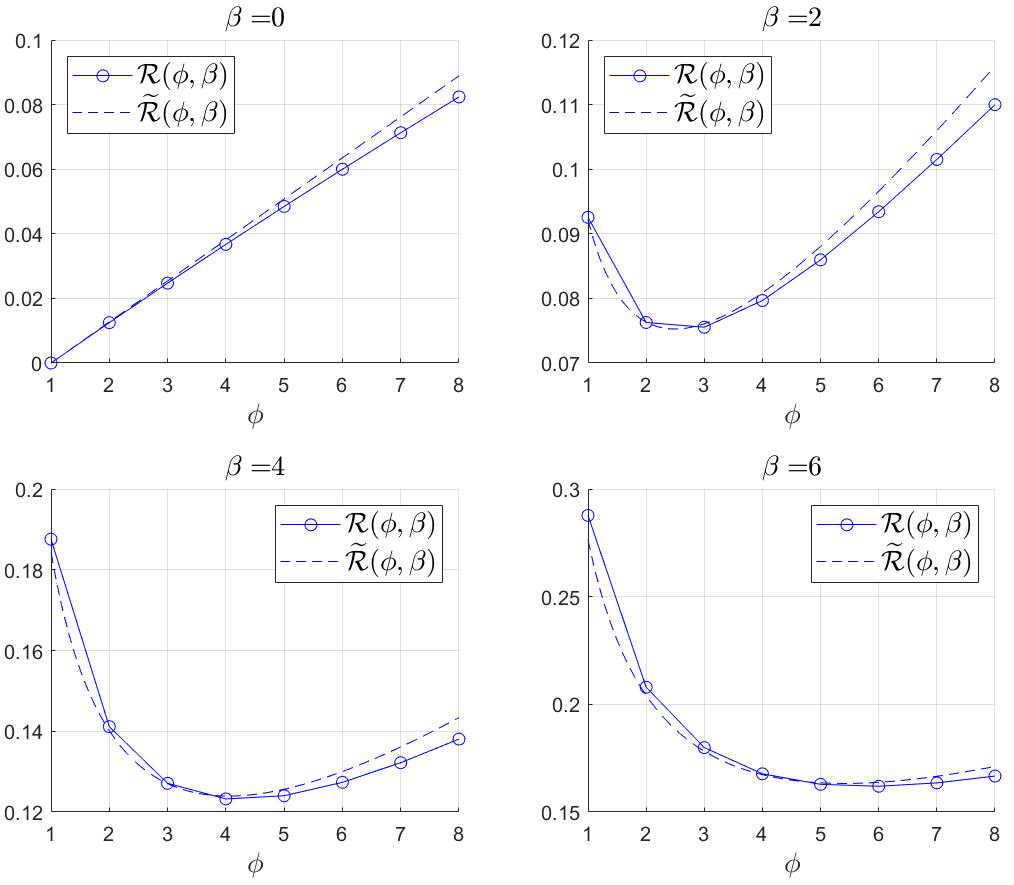}
	\caption{\small The graphs plot the personalization measure $\mathcal{R}(\phi,\beta)$ and its approximation $\widetilde{\mathcal{R}}(\phi,\beta)$ as a function of the time $\phi$ between consecutive interactions. Each panel corresponds to a different value of the parameter $\beta$. We consider the same market parameters and investment horizon as in Figure \ref{fig_jumps2_beta}, with $Y_0=1$, $P=I_2$, and each time step in the model corresponding to one month. The parameters of the risk aversion model are $\eta_n=0$, $p_{\epsilon}=0.05$, and $\sigma_{\epsilon}=0.64$.}\label{gamma_diff_approx}
\end{figure} 

The above proposition confirms two intuitive {claims}. First, in the absence of behavioral bias{es}, i.e.,  $\beta=0$, it is optimal for the robo-advisor to interact with the client at all times. Second, if there are no idiosyncratic risk aversion shocks, i.e., $p_{\epsilon}=0$, then it is optimal to never interact. 
Most interestingly, if both $\beta>0$ and $p_{\epsilon}>0$, then it may be suboptimal to interact at all times. 
This is because a higher frequency of interaction comes at the {expense of increased behavioral bias} in the communicated risk aversion parameter. Furthermore, the signs of the derivatives in (\ref{deltas}) show that a larger value of $\beta$ increases the optimal time between consecutive interactions, while larger values of $p_{\epsilon}$ and $\sigma_{\epsilon}$ imply a greater variance of idiosyncratic risk aversion shocks and push down the optimal time between interaction. In the proof of the proposition, we show that
\begin{align*}  
\frac{p_{\epsilon}\sigma_{\epsilon}}{\beta\sigma_0} <  1 \quad\Longrightarrow\quad \frac{\partial\widetilde{\mathcal{R}}(\phi,\beta)}{\partial\phi}\Big\lvert_{\phi=1} < 0,
\end{align*}
which gives a sufficient condition for when it is optimal to refrain from interacting at all times.\footnote{This condition is also close to being necessary. See Eq.\ (\ref{phiCond}){ in Appendix} \ref{AppendixC}. }
In the above inequality, we compare the rate of change of the idiosyncratic risk-aversion component, as measured by $p_{\epsilon}\sigma_{\epsilon}$, with the amount of behavioral bias, quantified by the product $\beta\sigma_0$ of the client's sensitivity to market returns and the volatility of returns. If the latter is greater than the former, then the derivative of $\widetilde{\mathcal{R}}$ at $\phi=1$ is negative, i.e., it is suboptimal to interact at all times. 


To ensure a high level of personalization, the robo-advisor must construct a process $(\gamma_n)_{n\geq 0}$ which is as close as possible to the client's actual risk aversion process $(\gamma_n^C)_{n\geq 0}$. 
{A related measure of} personalization may thus be directly built on the proximity of the investment strategy corresponding to $(\gamma_n)_{n\geq 0}$ and the strategy corresponding to $(\gamma_n^C)_{n\geq 0}$. The latter is the strategy that achieves full personalization and can only be attained if interaction occurs at all times and there {is} no behavioral bias, in which case the two risk aversion processes coincide.


Denote by $(\tilde\pi_n^*(\gamma_n))_{0\leq n<T}$ and $(\tilde\pi^*_n(\gamma_n^C))_{0\leq n<T}$ the optimal allocations corresponding to $(\gamma_n)_{0\leq n<T}$ and $(\gamma_n^C)_{0\leq n<T}$, respectively. In line with the definition of $\mathcal{R}(\phi,\beta)$ in (\ref{R}), we introduce the measure
\begin{align*}
\mathcal{S}(\phi,\beta):=\bbe\left[\frac{1}{T}\sum_{n=0}^{T-1}\Big|\frac{\tilde\pi_n^*(\gamma_n)-\tilde\pi_n^*(\gamma_n^C)}{\tilde\pi_n^*(\gamma_n^C)}\Big|\right].
\end{align*}
A smaller value of $\mathcal{S}(\phi,\beta)$ implies a higher level of personalization, and full personalization is achieved if and only if $\mathcal{S}(\phi,\beta)=0$. 
We observe that the difference between $\tilde\pi_n^*(\gamma_n)$ and $\tilde\pi_n^*(\gamma_n^C)$ is 
not a simple function of the difference between $\gamma_n$ and $\gamma_n^C$. Nevertheless, in Appendix \ref{AppendixC} we establish a direct relation between the two measures $\mathcal{R}$ and $\mathcal{S}$, namely,
\begin{align}\label{SRapprox}
\mathcal{S}(\phi,\beta) &= \mathcal{R}(\phi,\beta) +O((\phi-1)p_{\epsilon}\sigma_{\epsilon}^2) + O\Big(\frac{\beta^2\sigma^2}{\phi^2}\Big),
\end{align}
which shows that a small difference between $(\gamma_n)_{0\leq n<T}$ and $(\gamma_n^C)_{0\leq n<T}$ translates into a small difference between the corresponding investment strategies. In the above equation, the first error term is due to the idiosyncratic component of the client's risk aversion and it vanishes if $\phi=1$. The second error term originates from the client's behavioral bias and it is maximized at $\phi=1$.  

{Figure} \ref{RS_comparison} indicates that the value of $\phi$ that minimizes $\mathcal{R}(\phi,\beta)$ is a lower bound for the value of $\phi$ that minimizes $\mathcal{S}(\phi,\beta)$. In other words, minimizing $\mathcal{R}(\phi,\beta)$ provides a conservative estimate for the time between interaction that minimizes the allocation difference $\mathcal{S}(\phi,\beta)$.
{This is because $\mathcal{R}(\phi,\beta)$ underestimates the impact of the client's behavior bias. {Specifically,} $\mathcal{R}(\phi,\beta)$ equals the average relative difference between $(\gamma_n^C)_{n\geq 0}$ and $(\gamma_n)_{n\geq 0}$} in any single time interval defined by two consecutive interaction times. 
Instead, the optimal allocation at any given time depends on all future allocations and, thus, on the future path of the client's risk aversion. As a result, the behavioral bias in future investment decisions 
feeds into the investment decisions made at earlier times, and this is accounted for by the measure $\mathcal{S}(\phi,\beta)$. 

\begin{rem}
	Uniformly spaced interaction times yield a conservative estimate for the optimal {interaction} frequency. If the client were to interact with the robo-advisor at times triggered by market conditions, then we expect interaction to be more likely to occur after a period of extreme returns. The client would either be overly exuberant following a period of positive returns, or overly pessimistic after a period of negative returns. This magnifies the impact of the client's behavioral bias, 
	relative to a uniform interaction schedule, and would result in a larger value for $\phi$. \hfill\qed
\end{rem}

\begin{figure}[!ht]
	\centering
	\includegraphics[width=0.8\textwidth]{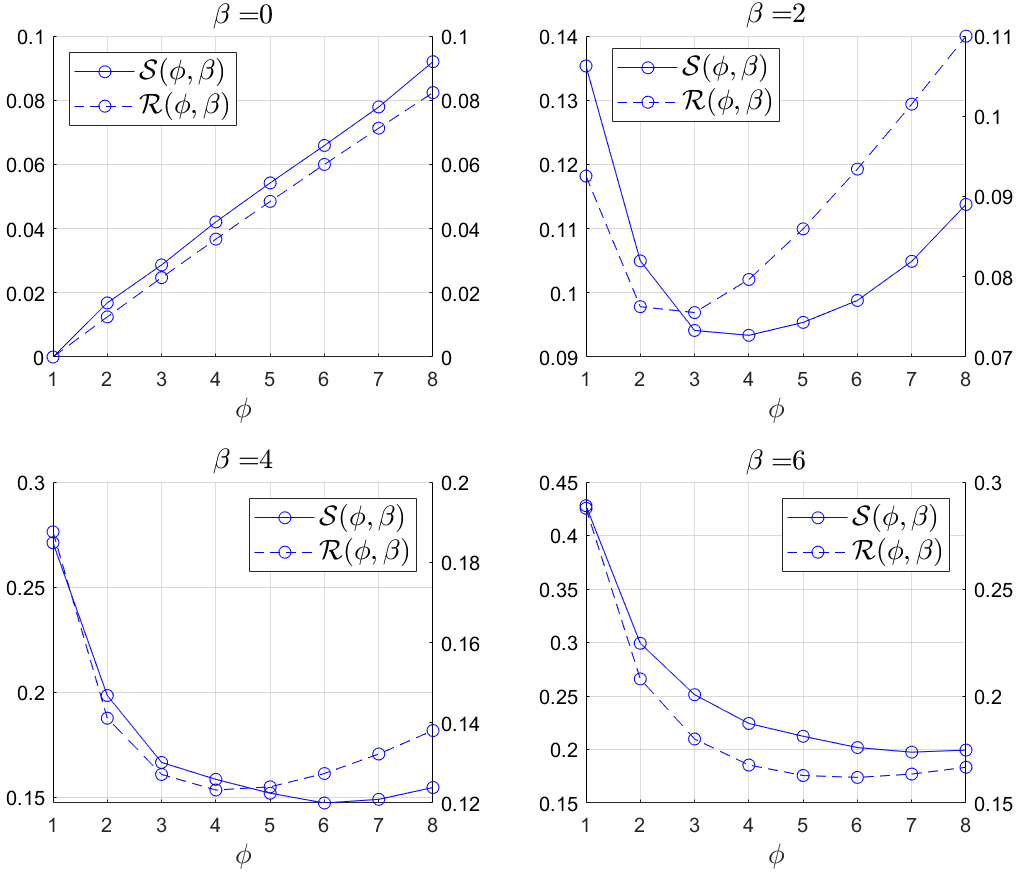}
	\caption{\small The graphs plot the personalization measures $\mathcal{S}(\phi,\beta)$ (left $y$-axis) and $\mathcal{R}(\phi,\beta)$ (right $y$-axis) as a function of the time $\phi$ between consecutive interactions. Each panel corresponds to a different value of the parameter $\beta$. The model parameters are the same as in Figure \ref{gamma_diff_approx}. 
	}\label{RS_comparison}
\end{figure} 

\subsection{Economic Transitions,  Risk Aversion, and Investment Performance}\label{SectionY}

We analyze how the optimal investment strategy is affected by economic state transitions and the corresponding changes in the client's risk aversion. To focus exclusively on the implications of economic transitions on investment decisions, we let the client's risk aversion depend only on the current economic state. {In this case, the} risk aversion {process} in (\ref{gammaEx}) takes the form $\gamma_n = \bar\gamma_n(Y_n)$. 
{We calibrate this process} so that the resulting optimal allocation $(\tilde\pi^*_n)_{0\leq n<T}$ is time-homogeneous, given the current economic state. Specifically, in Appendix \ref{AppendixC} (see Proposition \ref{PropGamma1}) we show that, for a given $\bar\pi>0$ and $\delta> -1$, 
there exists a unique risk aversion process {$(\gamma_n^{(\delta)})_{n\geq 0}$} such that the corresponding optimal allocation is given by
\begin{align}\label{timeHomogPi}
\tilde\pi^*_n = \left\{\begin{array}{ll} \bar\pi, \;& Y_n=1, \\
\bar\pi(1+\delta), \; &Y_n=2. \end{array}\right.
\end{align}
That is, in times of growth ($y=1$), the client's risky asset allocation is equal to $\bar\pi$,\footnote{For example, $\bar\pi=0.60$ corresponds to the classical 60/40 portfolio composition. This strategy was popularized by Jack Bogle, the founder of Vanguard, and is commonly used as a benchmark in portfolio allocation.} and when the economy transits to the recessionary regime ($y=2$), the allocation changes to $\bar\pi(1+\delta)$, where the value of $\delta$ {determines} the change in allocation. 

{It is well known from empirical studies that both the market Sharpe ratio and the risk aversion of retail investors are higher during periods of contractions. Consequently, retail investors may shift wealth away from the risky asset precisely when the benefit of investing, as measured by the expected reward per unit risk, is higher}.\footnote{The economic state variable $(Y_n)_{n\geq 0}$ is assumed to be observable, and it follows from (\ref{gammaCex}) that the client's risk aversion reacts instantaneously to a change in economic conditions. In reality, the economic state may be hidden, and its value can only be inferred probabilistically. We expect that extending the model to such a setting would not have a qualitative effect on our results. {Namely, the} extended model would capture 
the fact that, throughout the business cycle, the client is on average less willing to invest in the risky asset when the market Sharpe ratio is high.} For these investors, a robo-advisor which {\it {caters to the client's wishes}} {would construct a risk aversion process $(\gamma_n^{(\delta)})_{n\geq 0}$ such that $\delta<0$. By contrast, a robo-advisor which} {\it {goes against the client's wishes}} {would construct a risk aversion process $(\gamma_n^{(\delta)})_{n\geq 0}$ such that $\delta\geq 0$, i.e., invest more in the risky asset when the economy is in a state of contraction and the return per unit risk is high.}

The remainder of the section proceeds as follows. In Section~\ref{sec:Sharperatio}, we analyze the Sharpe ratio achieved by the optimal investment strategy and its dependence on the parameter $\delta$. In Section~\ref{sec:wealthdistr}, we analyze numerically the dependence of the client's terminal wealth distribution on $\delta$ and discuss how the robo-advisor should balance the client's risk {preferences} with the opportunity {to improve the portfolio's investment performance}.

\subsubsection{Sharpe Ratio of Optimal Investment Strategy}\label{sec:Sharperatio}

The Sharpe ratio of the strategy (\ref{timeHomogPi}) is defined as 
\begin{align}\label{sDef}
s^{\pi^*}(\delta) := \frac{\tilde\mu^{\pi^*}}{\sigma^{\pi^*}},
\end{align}
where $\tilde\mu^{\pi^*}$ and $\sigma^{\pi^*}$ are the long-run\footnote{{We compute the ``long-run'' Sharpe ratio of the strategy $\pi^*$, which is independent of the initial economic state, $Y_0$.}} mean and volatility of the excess returns {achieved by} $\pi^*$:
\begin{align*}
\tilde\mu^{\pi^*} &:= \lim_{T\to\infty}\bbe\Big[\frac{1}{T}\sum_{n=0}^{T-1}(r_{n,n+1}^{\pi^*}-r_{n+1})\Big], \quad
(\sigma^{\pi^*})^2 := {\lim_{T\to\infty}\bbe\Big[\frac{1}{T}\sum_{n=0}^{T-1}(r_{n,n+1}^{\pi^*}-r_{n+1}-\tilde\mu^{\pi^*})^2\Big]}.
\end{align*} 
We recall that $r_{n,n+1}^{\pi^*}$ is the return of {the strategy} $\pi^*$, between times $n$ and $n+1$, and that $r_{n+1}$ is the risk-free rate.  In Lemma \ref{LemSharpe} we show that the Sharpe ratio (\ref{sDef}) can be explicitly computed. 

Figure \ref{fig_sharpe} indicates that, for a fixed value of $\delta$, the portfolio's Sharpe ratio is increasing in the stationary probability $\lambda$ of the recessionary state ($y=2$). {Formally, $\lambda:=\lim_{n\to\infty}\bbp(Y_n=2)$, and $\lambda$ also equals the long-run proportion of time spent in the recessionary state.} {Additionally, the Sharpe ratio is} increasing in $a:=\tilde\mu(2)/\tilde\mu(1)$, which is the relative change in mean excess returns between the two states, and decreasing in $b:=\sigma(2)/\sigma(1)$, which is the relative change in the return volatility. These monotonicity properties turn out to hold under mild conditions (see Lemma \ref{lemDeriv}). 


Furthermore, Figure \ref{fig_sharpe} shows that the 
Sharpe ratio (\ref{sharpe}) is generally increasing in $\delta$, i.e., it is higher if a larger amount of wealth is invested in the risky asset when the economy is in a state of recession and, thus, the market Sharpe ratio is high. However, as the third panel shows, there may be situations when a {larger} value of $\delta$ may result in a smaller Sharpe ratio. 

In Lemma \ref{lemDeriv}, we derive the following characterization for the monotonicity of the Sharpe ratio with respect to $\delta$, 
\begin{align}\label{sharpe_pi}
\frac{\partial s^{\pi^*}(\delta)}{\partial \delta} > 0 \quad\Longleftrightarrow\quad 1+\frac{\sigma^2(1)}{\tilde\mu^2(1)} > a\Big(1+\frac{b^2\sigma^2(1)}{a^2\tilde\mu^2(1)}\Big)(1+\delta). 
\end{align}
Whether or not the condition above is satisfied depends on the values of $a$ and $b$, which determine the difference in market Sharpe ratios between the two economic states. In particular, consider the case where the allocation is the same in both economic states, i.e., $\delta=0$. For a fixed $b>1$, {it can be observed} from (\ref{sharpe_pi}) that there exists a threshold $a^*(b)>b$ that $a$ needs to exceed for the condition to be satisfied. 
In other words, for the portfolio's Sharpe ratio to increase with additional investment during recessions, the market Sharpe ratio needs to be sufficiently high in such an economic state. 
{The reason for why a marginally higher market Sharpe ratio may be insufficient is that larger market returns in an economic state increase both the numerator and the denominator of the Sharpe ratio, because the ``between-state'' standard deviation of returns is then higher.} We refer to the discussion following Lemma \ref{lemDeriv} 
for {additional} details.

Condition (\ref{sharpe_pi}) is generally {satisfied for empirically plausible parameter values}. That is, the portfolio's Sharpe ratio increases if more wealth is allocated to the risky asset when the market Sharpe ratio is high. However, Figure \ref{fig_sharpe} indicates that the gain is modest, and that the Sharpe ratio decreases more when the allocation to the risky asset is decreased, compared to how much it increases when the allocation to the risky asset is increased by the same amount.
In Appendix \ref{AppendixC}, we show that $s^{\pi^*}(\delta)$ is concave in a neighborhood of $\delta=0$, namely,
\begin{align}\label{pi_concave}
\left.\frac{\partial^2 s^{\pi^*}(\delta) }{\partial \delta^2}\right\lvert_{\delta=0}  &\propto -\frac{\sigma^2(1)}{\tilde\mu^2(1)}\Big(a^2+b^2\frac{\sigma^2(1)}{\tilde\mu^2(1)}\Big)\lambda + O(\lambda^2).
\end{align}
This confirms that in order to maintain a satisfactory Sharpe ratio, it is more important not to reduce the risky asset allocation in a state of recession  than to increase it.


\begin{figure}
\centering
\includegraphics[width=\textwidth]{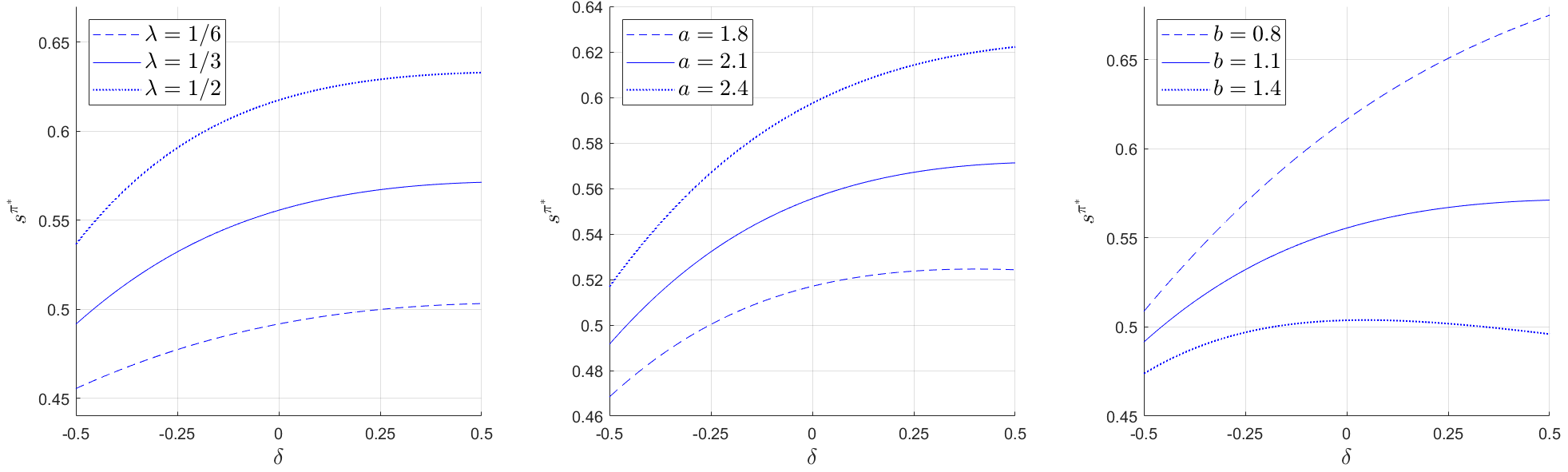}
\caption{\small The graphs plot the Sharpe ratio (\ref{sDef}) as a function of $\delta$. We set the state-dependent parameters $r,\mu,\sigma$ and the transition matrix $P$ to the values specified in (\ref{Lambda})-(\ref{musig}). This specification yields $\lambda=1/3$, $a=2.1$, and $b=1.1$. Left panel: We vary the parameter $\lambda$, while keeping $a$ and $b$ fixed. Middle panel: We vary the parameter $a$, while keeping $\lambda$ and $b$ fixed. Right panel: We vary the parameter $b$, while keeping $\lambda$ and $a$ fixed. Each time step in the model is one month, and we annualize the Sharpe ratio by multiplying (\ref{sDef}) by $\sqrt{12}$.
}\label{fig_sharpe} 
\end{figure}

\subsubsection{Wealth Distribution and Cyclicality of Risk Aversion}\label{sec:wealthdistr}

The portfolio's Sharpe ratio is determined by the first two moments of its returns, which may not be sufficient to characterize the entire return distribution. Therefore, comparing portfolios in terms of their Sharpe ratios therefore leaves out the impact of higher return moments. This is especially relevant in the presence of multiple economic states, given that the unconditional return distribution is then leptokurtic and skewed. 

We use numerical simulations to estimate the terminal distribution of the client's wealth, for different values of the parameter $\delta$. {We assume monthly portfolio rebalancing (a time step of one month) and set the transition matrix of the state variable $(Y_n)_{n\geq 0}$ to 
	\begin{align}\label{Lambda}   
P = \Bigg[\begin{array}{ll} P_{11} &\; P_{12} \\ P_{21} &\; P_{22} \end{array}\Bigg] = \Bigg[\begin{array}{ll} 0.95 &\; 0.05 \\ 0.10 &\; 0.90 \end{array}\Bigg].
\end{align}
{These transition probabilities are based on empirical values reported in \cite{Chauvet}, who also show that a regime switching model captures well the economic state transitions described by the NBER's business cycle chronology. These values show that, on average, economic expansions last longer than contractions.} 

We set the  state-dependent return parameters in accordance with \cite{TanWhitelaw}, who report historical averages for changes in the mean and volatility of stock market returns, as well as for changes in the market Sharpe ratio, between the peak of the business cycle and the subsequent trough. Specifically, we set the annual risk-free rate, mean and volatility of market returns to 
\begin{align}\label{musig}
r = (0.015,0), \qquad \mu=(0.081,0.137), \qquad \sigma=(0.155,0.173).
\end{align}  

{Figure} \ref{fig_sim_Y} shows the simulated distribution of the return of the optimal investment strategy {at maturity}, and the corresponding annualized rate of return. {It is evident from the figure that the skewness and kurtosis are higher when the risky asset allocation is} maintained or increased in a state of recession, and that in those cases, the return upside is considerably higher, with a limited additional downside risk. The same {pattern} is observed from the summary statistics {of} the simulations, reported in Table \ref{table1}. 



The above findings indicate that both the Sharpe ratio and the terminal wealth achieved by the strategy defined in (\ref{timeHomogPi}) are higher if $\delta\geq 0$, compared to the case $\delta<0$. In other words, the investment benefits are greater if the risky asset allocation is maintained or increased during recessions ($\delta\geq 0$). The question thus arises of \emph{how far the robo-advisor can reach ``against the will'' of a client} whose risk preferences are countercylical and thus consistent with the case $\delta<0$. 

While higher expected returns will be obtained in the long run, the client may suffer from adverse market moves in the short run, and may not have an adequate understanding of the long-term benefits. In a state of recession, with worsening economic outlook and risk aversion rising, the client may be particularly sensitive to what can be perceived as an investment mistake of the robo-advisor. {The importance of this dilemma faced by robo-advisors is emphasized by} \cite{Rossipeace}, who show empirically that algorithm aversion, i.e., the tendency of individuals to prefer a human forecaster over an algorithm, and to more quickly lose confidence in an algorithm than {in} a human after observing them make the same mistake (\cite{Simmons}), is one of the main obstacles for the {adoption of} robo-advising.

Our analysis suggests that a good middle ground for the robo-advisor is to encourage the client to simply maintain a fixed portfolio composition, which is consistent with the long-term investing principle of riding out business cycles. To that end, the robo-advisor may present statistics such as those reported in Table \ref{table1} and Figure \ref{fig_sim_Y} to the client. Thus, the client may observe that, over time, higher returns can be earned with limited additional risk by maintaining exposure to the risky asset during recessions instead of reducing it.\footnote{{In the context our model, this amounts to ``changing'' the client's risk aversion process from $(\gamma^{(\delta)}_n)_{n\geq 0}$ such that $\delta<0$, to $(\gamma^{(0)}_n)_{n\geq 0}$. We recall that $(\gamma_n^{(\delta)})$ is the risk aversion process implied by the strategy given in} (\ref{timeHomogPi}).}
This behavior would also be consistent with that of most robo-advising firms.
Namely, modifying the portfolio allocation based on economic conditions 
is a form of \emph{active management} and, in general, robo-advising firms {which focus} on long-term investing do not engage in such market timing.\footnote{One of the claimed benefits of robo-advising is that by managing the portfolio on the client's behalf, the client is helped to resist the temptation of attempting to time the market. The robo-advising firms Betterment and Wealthfront also mention empirical work, such as the Dalbar's annual \emph{Quantitative Analysis of Investors} report, which shows that investors who try to time the market tend to perform much worse than a ``buy-and-hold'' investor, and that the average investor significantly underperforms a broad stock market index. 
}
Rather, they urge clients to stay the course through changing market conditions, in order to reap the benefits of long-term investing. 

\begin{figure}[!ht]
\centering
\includegraphics[width=0.8\textwidth]{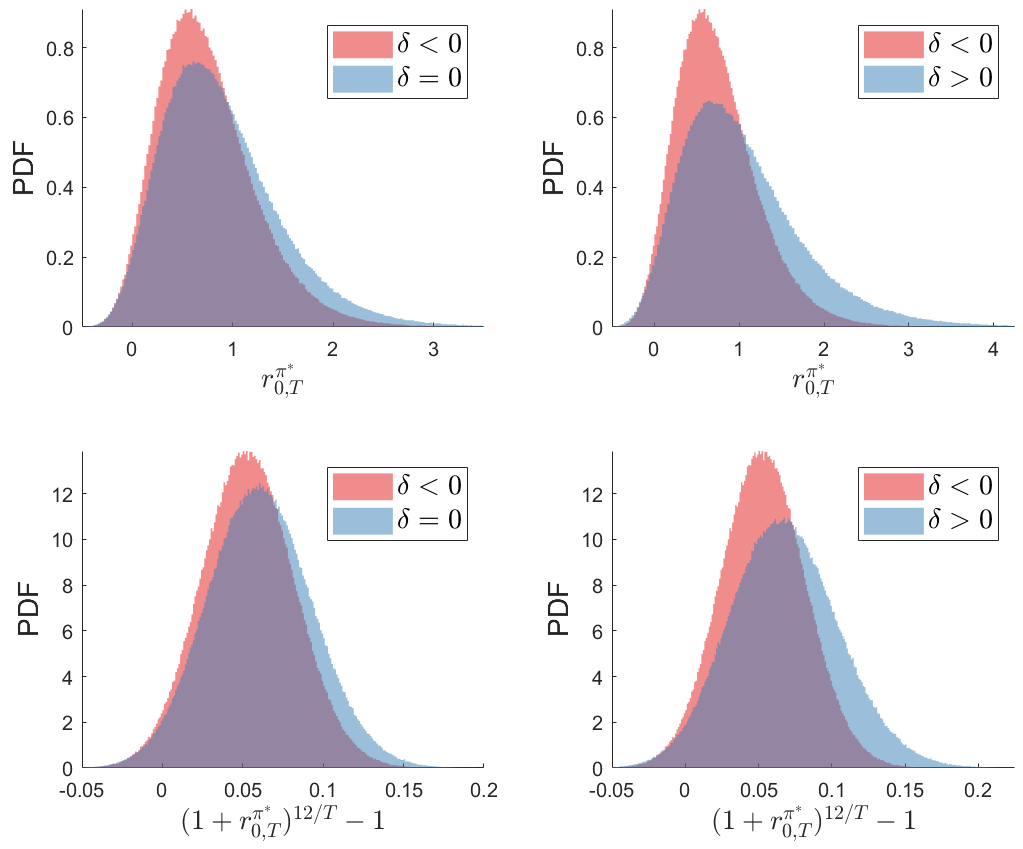}
\caption{\small Simulated distributions of the return of the strategy (\ref{timeHomogPi}) between time $0$ and time $T$ (top panels), and the corresponding annualized rate of return (bottom panels). The parameters of the strategy are $\bar\pi=0.6$ and $\delta\in\{-0.3,0,0.3\}$.
	One time step in the model corresponds to one month, the market parameters are given in (\ref{Lambda})-(\ref{musig}), the initial economic state is $Y_0=1$, and the investment horizon is $T=120$ months.}\label{fig_sim_Y}
\end{figure} 

\renewcommand*{\arraystretch}{1.3}
\begin{table}[h!]
\begin{center}
	{
	\begin{tabular}{|c|c|c|c|c|c|c|c|} \hline 
		& Mean & SD & Skewness & Kurtosis & 90\% VaR & 95\% VaR  & 99\% VaR \\ \hline
		$\delta<0$	& 0.740 & 0.485  & 0.838 & 4.257 & -0.179 & -0.067 & 0.117  \\
		$\delta=0$	& 0.881 & 0.589  & 0.970 & 4.712 & -0.213 & -0.085 & 0.118 \\
		$\delta>0$	& 1.036 & 0.735  & 1.231 & 5.934 & -0.233 & -0.091 & 0.134  \\ \hline 
		$\delta<0$	& 0.053 & 0.029  & 0.066 & 3.011 & -0.017 & -0.006 & 0.012  \\
		$\delta=0$	& 0.061 & 0.032  & 0.092 & 3.016 & -0.019 & -0.008 & 0.012 \\ 
		$\delta>0$	& 0.068 & 0.037  & 0.162 & 3.089 & -0.021 & -0.009 & 0.014  \\ \hline
	\end{tabular}}
\end{center}
\vspace{-10pt}
\caption{\small Summary statistics for the simulated return distributions in Figure \ref{fig_sim_Y}. The top half of the table reports statistics for the return of the strategy throughout the investment {period}. The bottom half of the table gives the statistics for the annualized rate of return.}\label{table1}
\end{table}

\section{Concluding Remarks}\label{sec:conclusions}

The past decade has witnessed the rapid emergence of robo-advisors, which are 
{online platforms allowing clients to interact directly with automated investment algorithms}. Recent literature has provided empirical evidence on the characteristics of those investors who are more likely to switch to robo-advising, and how robo-advising tools affect the portfolio composition of investors.

In {this} work, we build a novel framework {that incorporates} important features of modern robo-advising systems. 
In this framework, a robo-advisor undertakes the task of optimally investing a client's wealth. 
The client has a risk profile that changes {in response to} market returns, economic conditions, and idiosyncratic events, and to which the robo-advisor's investment performance criterion dynamically adapts. 
We derive and analyze optimal investment strategies, and highlight the role of risk aversion process in driving the intertemporal hedging demand.


The frequency of interaction between the client and robo-advisor determines the level of portfolio personalization, which is defined in terms of how accurately the robo-advisor is able to track the client's risk profile. We show that there exists an optimal interaction frequency, which maximizes portfolio personalization by striking a balance between receiving information from the client in a timely manner, and mitigating the effect of the client's behavioral biases. 

We quantify how the Sharpe ratio of the optimal investment strategy depends on the allocation of wealth within economic regimes, and study the potential gains achieved by a robo-advisor's strategy  during periods of high market Sharpe ratio. The implementation of this strategy may require the robo-advisor to go against the client's wishes, 
because both the client's risk aversion and the market Sharpe ratio tend to be countercyclical. We show that by simply rebalancing the portfolio to maintain constant weights throughout the business cycle, the portfolio's Sharpe ratio is close to being optimal. We also show that the portfolio return distribution is significantly improved {relative to the situation where the investment in the risky asset is reduced during periods of economic contractions}.

Our framework can be extended and further aligned to the practices of modern robo-advising systems. 
Indeed, the market model can be allowed to include 
 multiple tradable assets 
{and portfolio rebalanced only if the portfolio weights have drifted significantly from the target weights, due to price moves of the underlying securities. Robo-advisors generally use such threshold updating rules to rebalance their portfolios (\cite{Beketov}), while minimizing expense ratios and maximizing tax efficiency. 
Furthermore, the regime switching model can be generalized to make the economic state variable hidden, in line with the fact that economic conditions are typically not {transparently} observed, or observed with a lag.} 
If the state variable is hidden, then the robo-advisor {runs the risk of not} acting {in the client's} best interest {by making investment decisions} based on misclassified economic conditions.  
{Finally, one may account for noise in the information communicated by the client. {As} the level of noise {is} expected to increase with the frequency of interaction, this would act as a further incentive to reduce interaction with the client.} 


Herein, market returns are assumed to be conditionally independent of economic state transitions. Nevertheless, these transitions are commonly associated with significant market moves, in particular when the economy enters a recession.\footnote{This is also the case in the consumption-based asset pricing frameworks which generate  countercyclical variation in risk aversion, such as \cite{Campbell}. In their framework, a worsening economic outlook leads to an increase in risk aversion, and investors demand higher reward for carrying risk, which in turn leads to a fall in stock prices.} Such market moves would effectively result in the client ``buying high'' and ``selling low'', 
and thus exacerbate the adverse effects of comonotonicity between risk aversion and the market Sharpe ratio. 
{This would further incentivize the robo-advisor to act against the client's wishes, e.g., if a market drop is likely to happen \emph{following} a change in economic conditions, the portfolio's market exposure could be adjusted to account for the probability of a near term market drop.}\footnote{Portfolio allocation under regime switching is widely studied. Related to our work, \cite{TanWhitelaw} show how simple market timing strategies can exploit the fact that Sharpe ratios are countercyclical, and \cite{Coudert} show that risk aversion indices published by financial institutions are good leading indicators of stock market crises. \cite{Kritzman} show how a Markov switching model can be used to avoid large portfolio losses related to ``event regimes'', during which asset {return} dynamics start deviating from their past levels and the market risk premium is low.}

\appendix

\section{Properties of Optimal Investment Strategy}\label{Appendix0}

\subsection{Computation of Optimal Investment Strategy}\label{A0part1}

In the following proposition, {we provide a characterization of the optimal investment strategy given in Proposition} \ref{prop1}, {which depends on recursively computable expressions.} 


\begin{prop}\label{PropUhat}
	The optimal allocation in Proposition \ref{prop1} {admits the representation}
	\begin{align}\label{uhatprop2}
	\tilde\pi^*_n(d) 
	&= \frac{1}{\gamma_n}\frac{\mu_n^{az}(d)-R_{n+1}\gamma_n(\mu_n^{bz}(d)
		-\mu_n^{a}(d)\mu_n^{az}(d))}{\mu_n^{bz^2}(d)- (\mu_n^{az}(d))^2}, 
	\end{align}
	for any $d\in\mathcal{D}_n$, where 
	\begin{align}\label{muamub}
	\begin{split} 
	\begin{array}{llll}
	\mu_n^{a}(d) := \bbe_{n,d}[a_{n+1}(D_{n+1})], &\;
	\mu_n^{az}(d) := \bbe_{n,d}[a_{n+1}(D_{n+1})\widetilde Z_{n+1}],& \\
	\mu_n^{b}(d) := \bbe_{n,d}[b_{n+1}(D_{n+1})], &\; 
	\mu_n^{bz}(d) := \bbe_{n,d}[b_{n+1}(D_{n+1})\widetilde Z_{n+1}], &\; 
	\mu_n^{bz^2}(d) := \bbe_{n,d}[b_{n+1}(D_{n+1})\widetilde Z_{n+1}^2].
	\end{array}
	\end{split}
	\end{align}
	The sequences $(a_n(d))_{0\leq n<T}$ and $(b_n(d))_{0\leq n<T}$ {satisfy the recursions}
	\begin{align}\label{abRecursions} 
	\begin{split}
	a_{n}(d) &:=\bbe_{n,d}[(R_{n+1}+\widetilde Z_{n+1}\tilde\pi_{n}^*(d))a_{n+1}(D_{n+1})], \\
	b_{n}(d)  &:=\bbe_{n,d}[(R_{n+1}+\widetilde Z_{n+1}\tilde\pi_{n}^*(d))^2b_{n+1}(D_{n+1})],  
	\end{split}
	\end{align}
	with $a_{T}(d)=b_{T}(d)=1$, for all $d\in\mathcal{D}_T$. 
\end{prop}

The {quantities} $a_n(d)$ and $b_n(d)$ in Proposition \ref{PropUhat} are the first and second moments of the future value of one dollar invested optimally between time $n$ and the {terminal} date $T$, conditionally on $D_n=d$. That is, 
\begin{align}\label{abr}
a_{n}(d) &:= \bbe_{n,d}[1+r_{n,T}^{\pi^*}],\qquad 
b_{n}(d) := \bbe_{n,d}[(1+r_{n,T}^{\pi^*})^2], 
\end{align}
with the simple return $r_{n,T}^{\pi^*}$ defined in (\ref{return}). 
{Because both $a_{n}(d)$ and $b_{n}(d)$} are independent of the wealth $x$ {at time $n$}, it follows from (\ref{V}) that the value function at time $n$ is also independent of $x$, and given by
\begin{align}\label{optimalV}
V_{n}(x,d) = V_n(d) := a_{n}(d)-1-\frac{\gamma_n(d)}{2}(b_{n}(d)-a_{n}^2(d)). 
\end{align}
Finally, {Proposition} \ref{PropUhat} shows that the optimal proportion of wealth allocated to the risky asset, $\tilde\pi^*_n$, is also independent of the wealth variable.

The expected values in (\ref{muamub}) admit integral representations, which in turn can be used to compute the expected values in (\ref{abRecursions}). Next, we provide such a representation for $\mu_n^{az}(d)$, both in the general case and in the specific case described in Section \ref{sectionModel}. 
\begin{lem}\label{LemMu}
	\begin{itemize}
		\item[(a)]  Assume that given $D_n=d$, the random variable $\epsilon_{(n+1)}$ admits a generalized probability density function, $f_{\epsilon_{(n+1)}|d}$. Then,
		\begin{align*}
		\mu_n^{az}(d) &= \sum_{y'\in\mathcal{Y}}P_{y_n,y'}\int_{z'\in\bbr}\int_{\epsilon'\in\bbr^{n+1}}a_{n+1}(D_{n+1})\tilde z'f_{Z|y_n}(z')f_{\epsilon_{(n+1)}|d}(\epsilon')dz'd\epsilon',
		\end{align*}
		where $\tilde z':=z'-r(y_n)$, and $D_{n+1}=(Y_{(n+1)},Z_{(n+1)},\tau_{(n+1)},\xi_{(n+1)})\in\mathcal{D}_{n+1}$ is such that $Y_{(n+1)}=(y,y')$, $Z_{(n+1)}=(z,z')$, $\tau_{(n+1)}=(\tau,\tau_{n+1})$, and $\xi_{(n+1)}=(\xi,\xi_{n+1})$. The value of $\tau_{n+1}$ is determined by the triplet $(Y_{(n+1)},Z_{(n+1)},\epsilon_{(n+1)})$. If $\tau_{n+1}=\tau_n$, then $\xi_{n+1}=\xi_n$, while if $\tau_{n+1}=n+1$,  $\xi_{n+1}$ is determined by $(Y_{(n+1)},Z_{(n+1)},\epsilon_{(n+1)})$.
		\item[(b)]
		For the model in Section \ref{sectionModel}, i.e., where the risk aversion process $(\gamma_n)_{n\geq 0}$ is given by (\ref{gammaEx}) and the interaction times are of the form $T_k=k\phi$, for some $\phi\geq 1$, we have 
		\begin{align}\label{mua_ex}
		\mu_n^{az}(d) &= \Bigg\{\begin{array}{ll} \sum_{y'\in\mathcal{Y}}P_{y_n,y'}\int_{z'\in\bbr}a_{n+1}(D_{n+1})\tilde z'f_{Z|y_n}(z')dz', &\; \tau_{n+1}<n+1, \\
		\sum_{y'\in\mathcal{Y}}P_{y_n,y'}\int_{z'\in\bbr}\int_{\epsilon'\in\bbr}a_{n+1}(D_{n+1})\tilde z'f_{Z|y_n}(z')f_{\epsilon}^{(\phi)}(\epsilon')dz'd\epsilon', &\; \tau_{n+1}=n+1,
		\end{array}\Bigg.
		\end{align}	
		where $f_{Z|y_n}$ is the Gaussian {probability} density {function} with mean $\mu(y_n)$ and variance $\sigma^2(y_n)$, $f_{\epsilon}^{(\phi)}$ is the $\phi$-fold convolution of the generalized density function of the i.i.d.\ sequence $(\epsilon_n)_{n\geq 1}$, and\footnote{With a deterministic interaction schedule, we omit the interaction times $\tau_n$ from the state variable $D_n$.}
		\begin{align*}
		D_{n+1} &= \Bigg\{\begin{array}{ll} \big((y,y'),(z,z'),(\xi,\xi_n)\big), &\; \tau_{n+1}<n+1, \\
		\big((y,y'),(z,z'),(\xi,\xi_{n+1})\big), &\; \tau_{n+1}=n+1.
		\end{array}\Bigg.
		\end{align*}	
		If $\tau_{n+1}=n+1$, then $\xi_{n+1}$ is given by an explicit function of $((y,y'),(z,z'),\xi_n,\epsilon')$. 
	\end{itemize}
\end{lem}

Equation \eqref{mua_ex} shows that two distinct cases need to be considered, depending on whether or not the client's risk preferences are solicited {at time $n$.} 
In both cases, the computation of $\mu_n^{az}(d)$ involves a sum over the distribution of the economic state $Y_{n+1}$, and an integral with respect to the distribution of the market return $Z_{n+1}$. 
Additionally, if $\tau_{n+1}= n+1$, the client's {risk preferences are solicited} at time $n+1$ and $\mu_n^{az}(d)$ involves an additional integral with respect to the distribution of the {cumulative} idiosyncratic risk aversion shock $\sum_{k=\tau_n}^{n}\epsilon_{k+1}$. 
Hence, the probability {distributions} $f_{Z|y_n}$, $P_{y_n,\cdot}$, and $f_{\epsilon}^{(\phi)}$, link the optimal allocation at time $n$ 
to the optimal allocations at time $n+1$. 
We observe that if $\tau_{n+1}<n+1$, the dynamics are only based on incoming market and economic information, modeled through $Z_{n+1}$ and $Y_{n+1}$, while, in the case $\tau_{n+1}=n+1$, they also depend on information received from the client, given by $\sum_{k=\tau_n}^{n}\epsilon_{k+1}$.  

\begin{rem}\label{remDist}
	{Equation} (\ref{abRecursions}) shows how the first two moments of $1+r_{n,T}^{\pi^*}$ can be computed recursively. This can be extended to the $m$-th finite moment, $\mu^{(m)}_{n}(d) := \bbe_{n,d}[(1+r_{n,T}^{\pi^*})^m]$, which satisfies the recursion
	\begin{align*}
	\mu^{(m)}_{n}(d)  &=\bbe_{n,d}\big[\big(R_{n+1}+\widetilde Z_{n+1}\tilde\pi_{n}^*(d)\big)^m\mu^{(m)}_{n+1}(D_{n+1})\big].
	\end{align*}
	Under mild conditions, a probability distribution with finite moments of all orders is uniquely determined by its moments (see \cite{Billingsley}, Sec.\ 30). In that case, 
	the distribution of the portfolio return $r_{n,T}^{\pi^*}$ can be recovered from a finite set of moments $\mu^{(1)}_n,\dots,\mu^{(M)}_n$, for some $M\geq 1$.
	\hfill\qed 
\end{rem}

\subsection{Computational Complexity of Optimal Investment Strategy
}\label{AppendixComplexity} 

The optimal investment strategy in Proposition \ref{PropUhat} is computed using backward induction, which requires discretization of the state variable $D_n$, {for each $n$}. This variable is high-dimensional, {as it contains} the history of economic states, market returns, interaction times, and communicated risk aversion parameters. 
{However, the optimal investment strategy at time $n$ can in general be written as a function of a lower-dimensional random variable $\widetilde{D}_n$, which is measurable with respect to $D_n$.} For instance, in the model of Section \ref{sectionModel},  the variable $\widetilde D_n$ consists of four components:\footnote{The functional $J_n$ in (\ref{BjorkCriterion}) depends on $d\in\mathcal{D}_n$. 
	In this example, 
	for a fixed $\tilde d\in \bbr_0^+\times \bbr^2\times\mathcal{Y}$, 
	the value of $J_n$ is the same for any $d\in\mathcal{D}_n$ such that if $D_n=d$ then $\widetilde D_n=\tilde d$.}
\begin{align*}
\widetilde D_n := \Big(\xi_{n},\sum_{k=\tau_n-\phi}^{\tau_n-1}(Z_{k+1}-\mu_{k+1}),\sum_{k=\tau_n}^{n-1}\big(Z_{k+1}-\mu_{k+1}\big),Y_n\Big) \in\bbr_0^+\times \bbr^2\times \mathcal{Y}.
\end{align*}
That is, $\widetilde D_n$ consists of the most recently communicated risk aversion parameter, the sum of excess market returns between the two most recent interaction times, the sum of excess market returns since the most recent interaction time, and the current economic state. 
If a step size $\Delta\gamma$ is used to discretize the risk aversion {parameter}, a step size $\Delta z$ is used for each of the return variables, and a {general} number $|\mathcal{Y}|\geq 1$ of economic states is assumed, the number of points in the discretization grid is of order
\begin{align*} 
\frac{T|\mathcal{Y}|}{\Delta\gamma\Delta z}\Big(\frac{1}{\Delta z}{\bf 1}_{\{\phi>1\}}+ {\bf 1}_{\{\phi=1\}}\Big),
\end{align*}
where each point requires numerical integration to compute each of the expected values in (\ref{muamub}) 
(see (\ref{mua_ex})). 




In the general case, one may use properties of the risk aversion process to reduce the computational complexity. 
For example, to compute $\mu_n^{az}$ in Lemma \ref{LemMu}, 
the numerical integration can be avoided by evaluating $Z_{n+1}$ at its expected value, $\mu_{n+1}$. 
This leads to the approximation
\begin{align*}
\mu_n^{az}(d) 
&= \bbe_{n,d}[a_{n+1}(D_{n+1})\tilde Z_{n+1}|Z_{n+1}=\mu_{n+1}] + O\Big(\Big(\sum_{k=n+1}^{T-1}\bbe\big[(\tilde\pi^*_k(\gamma_k)-\tilde\pi^*_k(\gamma'_k))^2\big]\Big)^{1/2}\Big).
\end{align*}
In the above error term, $(\tilde\pi^*(\gamma_k'))_{k=n+1}^{T-1}$ are the future optimal allocations corresponding to $(\gamma_k')_{k=n+1}^{T-1}$, {which is} the risk aversion process resulting from setting $Z_{n+1}=\mu_{n+1}$. In particular, if the future risk aversion values are independent of $Z_{n+1}$, then $(\gamma_k')_{k=n+1}^{T-1}=(\gamma_k)_{k=n+1}^{T-1}$, and the error term vanishes. In general, the effect of a single market return on the future risk aversion path is limited - either temporary, or quickly diluted by the effect of future returns. For instance, in the risk aversion model of Section \ref{sectionModel}, the return $Z_{n+1}$ impacts the risk aversion process only in a time interval bounded by two consecutive interaction times.  

The computational complexity can also be reduced by using properties of the economic state transition matrix. In particular, for a single time step, the transition probability out of the current state is generally small (see (\ref{Lambda})). In the computation of $\mu_n^{az}(d)$, this leads to the  approximation
\begin{align*}
\mu_n^{az}(d) 
= \sum_{y'\in\widetilde{\mathcal{Y}}}P_{y_n,y'}\bbe_{n,d}\big[a_{n+1}(D_{n+1})\tilde Z_{n+1}\big|Y_{n+1}=y'\big] + O\Big(1-\sum_{y'\in\widetilde{\mathcal{Y}}}P_{y_n,y'}\Big),
\end{align*}
where $\widetilde{\mathcal{Y}}\subseteq\mathcal{Y}$. For instance, {if $\widetilde{\mathcal{Y}}=\{y_n\}$, we condition on staying in the current economic state,} while if $\widetilde{\mathcal{Y}}=\{y_n-1,y_n,y_n+1\}$, {we condition on staying in the current state or transitioning to neighboring states.}

\subsection{Investment Constraints}\label{secConstraints}
{
	
	{The dynamic optimization problem analyzed in} Section~\ref{SectionModel} {can be enhanced to handle portfolio constraints {typically imposed} by robo-advising firms.} 
	For example, we may assume that the space of admissible {strategies} is restricted to control laws $\pi=(\pi_n)_{0\leq n<T}$ such that the proportion of wealth invested in the risky asset, denoted by $\tilde\pi=(\tilde\pi_n)_{0\leq n<T}$, satisfies
	\begin{align}\label{constraint}
	-\infty<\underline{\pi}_n\leq \tilde\pi_n \leq \bar\pi_n<\infty,
	\end{align} 
	where $(\underline{\pi}_n)_{0\leq n<T}$ and $(\overline{\pi}_n)_{0\leq n<T}$ are deterministic sequences. In particular, no short-selling and no borrowing constraints correspond to $\underline{\pi}\equiv 0$ and $\overline\pi\equiv 1$, respectively. 
	
	In this case, the optimal allocation formula is given by a truncated version of the unconstrained optimal allocation formula. Namely, in the unconstrained case, the proof of Proposition \ref{prop1} shows that the optimal allocation $\tilde\pi^*_n$ is obtained by maximizing a second order polynomial $h_n$ with a negative leading order coefficient (see (\ref{hDef})). 
	Under the above portfolio constraint, the optimal allocation is thus obtained by truncating the outcome of the unconstrained allocation formula (\ref{uhatprop2}) from below by $\underline{\pi}_n$, and from above by $\overline{\pi}_n$. 
	
}

{
	
	We next consider the effect of truncating {the} optimal investment strategy $\pi^*$ to zero whenever the wealth level is {negative}.} This results in the investment strategy $\pi^o$ given by
	\begin{align}\label{pi0}
	\pi_n^o(x,d) = \pi_n^*(x,d){\bf 1}_{\{x\geq 0\}}. 
	\end{align}
	Hence, $\pi^o$ coincides with the optimal strategy $\pi^*$ if the wealth variable $x$ is non-negative, {but switches} to a risk-free portfolio when the wealth becomes negative. In the following lemma, we study the difference between the strategies $\pi^o$ and $\pi^*$. 
	
	\begin{lem}\label{lem0} Let $\pi^*$ be the optimal strategy in Proposition \ref{prop1} under the constraint (\ref{constraint}) with $\underline{\pi}\equiv 0$ and $\bar\pi\equiv 1$. For $x>0$ and $d\in\mathcal{D}_n$, we have the following relations between the expectation and variance of the portfolio return achieved under the strategy $\pi^*$ and the constrained strategy $\pi^o$ defined in (\ref{pi0}):
		\begin{align*}
		\bbe_{n,x,d}[1+r_{n,T}^{\pi^o}] 
		&= \bbe_{n,d}[1+r_{n,T}^{\pi^*}] + O(\bbp_{n,d}(\widetilde Z_{n+1}<-1)), \\
		Var_{n,x,d}[1+r_{n,T}^{\pi^o}] &= Var_{n,d}[1+r_{n,T}^{\pi^{\pi^*}}] + O(\bbp_{n,d}(\widetilde Z_{n+1}<-1)). 
		\end{align*}
		Furthermore, for $x>0$ and $d\in\mathcal{D}_n$, the objective functional (\ref{BjorkCriterion}) evaluated at $\pi^o$ satisfies the relation
		\begin{align*}
		J_n(x,d,\pi^o) &= V_n(d) + O(\bbp_{n,d}(\widetilde Z_{n+1}<-1)),
		\end{align*}
		where $V_n(d)$ is the value function in (\ref{optimalV}). 
	\end{lem} 
	
	Observe that in the previous lemma, the expectation and variance terms under $\pi^o$ do not depend on the value of $x$, as long as $x>0$. {The same is true if $x<0$, in which case they only depend on the risk-free rate throughout the investment horizon, and can be written as}
	\begin{align*}
	\bbe_{n,x,d}[1+r_{n,T}^{\pi^o}] &= \bbe_{n,d}\Big[\prod_{k=n}^{T-1}R_{k+1}\Big], \qquad
	Var_{n,x,d}[1+r_{n,T}^{\pi^o}] = Var_{n,d}\Big[\prod_{k=n}^{T-1}R_{k+1}\Big]. 
	\end{align*}
	
	The proof of the lemma shows that {the} constants in the $\mathcal{O}$-terms admit explicit upper bounds depending only on $n$ and can thus be uniformly bounded for all $0\leq n<T$. The order of the 
	$\mathcal{O}$-terms can also be uniformly bounded by a constant independent of $n$ and $d$, because
	\begin{align*}
	\bbp_{n,d}(\widetilde Z_{n+1}<-1) \leq \max_{y\in\mathcal{Y}} \bbp(\widetilde Z_1<-1|Y_0=y).
	\end{align*}
	For reasonable parameter values, the above probability is very small. For example, standard annualized parameter values for the mean and variance of $\widetilde Z_{1}$ are $\tilde\mu=0.1$ and $\sigma=0.2$. {In that case, the event $\{Z_1<-1\}$ corresponds to a negative return whose magnitude exceeds five standard deviations.} As the discrete step size in the model gets smaller (e.g., monthly rather than annual), the probability becomes smaller, and vanishes in the continuous-time limit. 
	This can be seen explicitly for the model considered in Section \ref{sectionModel}, where returns are conditionally Gaussian. In that case, we have the bound
	\begin{align*}
	\bbp(\widetilde Z_{n+1}<-1|Y_n=y) \leq {\sigma(y)}e^{-\frac{1}{2\sigma^2(y)}}, 
	\end{align*}
	and the right-hand side converges to zero as $\sigma(y)$ tends to zero. 
}

\section{Proofs and Auxiliary Results for Section \ref{SectionOptimalAllocation}}\label{AppendixA}

We start by proving a lemma that will be used in the proof of Proposition \ref{prop1}. 

\begin{lem}\label{lem_ab}
	Let $\pi=(\pi_n)_{n\geq 0}$ be an admissible control law of the form 
	\begin{align}\label{pi_x}
	\pi_n(x,d) = \tilde\pi_{n}(d) x, 
	\end{align}
	for any {$x\in\bbr$}, and $d\in\mathcal{D}_n$. Then, the return moments 
	\begin{align*}
	a_n^{\pi}(d) := \bbe_{n,d}[1+r_{n,T}^{\pi}], 
\qquad
	b_{n}^{\pi}(d) := \bbe_{n,d}[(1+r_{n,T}^{\pi})^2], 
	\end{align*}
	satisfy the recursions in (\ref{abRecursions}) with $\pi^*=\pi$. 
\end{lem} 

\paragraph{Proof of Lemma \ref{lem_ab}:} 
For $n=T-1$, {it follows from} the wealth dynamics (\ref{dX}) that
\begin{align*} 
\bbe_{n,d}[1+r_{n,T}^{\pi}] &= \bbe_{n,d}[R_{n+1}+\widetilde Z_{n+1} \tilde\pi_{n}(d)] =R_{n+1}+\tilde\mu_{n+1} \tilde\pi_{n}(d).
\end{align*} 
Next, let $n\in\{0,1,\dots,T-2\}$ and assume that the result holds for $n+1,n+2,\dots,T-1$. Then,
\begin{align*} 
\bbe_{n,d}[1+r_{n,T}^{\pi}] 
&= \bbe_{n,x, d}\Big[(R_{n+1}+\widetilde Z_{n+1}\tilde\pi_{n}(d))\prod_{k=n+1}^{T-1}(R_{k+1}+\widetilde Z_{k+1}\tilde\pi_{k}(D_k))\Big]\\
&= \bbe_{n,x, d}\Big[(R_{n+1}+\widetilde Z_{n+1}\tilde\pi_{n}(d))\bbe_{n+1,X_{n+1}^{\pi}, D_{n+1}}\Big[\prod_{k=n+1}^{T-1}(R_{k+1}+\widetilde Z_{k+1}\tilde\pi_{k}(D_k))\Big]\Big]\\
&= \bbe_{n,x, d}[(R_{n+1}+\widetilde Z_{n+1}\tilde\pi_{n}(d))a_{n+1}^{\pi}( D_{n+1})] \\
&= \bbe_{n,d}[(R_{n+1}+\widetilde Z_{n+1}\tilde\pi_{n}(d))a_{n+1}^{\pi}( D_{n+1})],
\end{align*}
The assertion for the $b_n$-sequence is shown in the same way. 
\hfill\qed

\medskip Next, we derive 
the extended HJB system of equations necessarily satisfied by an optimal control law for the optimization problem defined by (\ref{BjorkCriterion})-(\ref{BjorkCriterion2}). To simplify notation, 
we introduce the random variable 
\begin{align*}
\widetilde D_n:=(X_{n},D_n)\in\widetilde{\mathcal{D}}_{n+1}:={\bbrZ}\times \mathcal{D}_n.
\end{align*} 
{Furthermore,} we write
\begin{align*}
\widetilde D_{n+1}^{\pi}=(X_{n+1}^{\pi},D_{n+1}), 
\end{align*}
where $X_{n+1}^{\pi}$ is obtained by applying the control law $\pi$ to the wealth $x$ at time $n$. 

\begin{prop}\label{PropHJB}
	Assume that an optimal control law $\pi^*$ for the optimization problem (\ref{BjorkCriterion})-(\ref{BjorkCriterion2}) exists. Then, the value function (\ref{V}) satisfies the  recursive equation 
	\begin{align}\label{HJB}
	V_{n}(\tilde d)
	=\sup_{\pi}\Big\{
	\bbe_{n,\tilde d}[V_{n+1}(\widetilde D^{\pi}_{n+1})]&-\Big(\bbe_{n,\tilde d}[f_{n+1,n+1}(\widetilde D_{n+1}^{\pi};\widetilde D_{n+1}^{\pi})] -\bbe_{n,\tilde d}[f_{n+1,n}(\widetilde D_{n+1}^{\pi};\tilde d)]\Big)\\
	&-\Big(\bbe_{n,\tilde d}\Big[\frac{\gamma_{n+1}(D_{n+1})}{2}\Big(\frac{g_{n+1}(\widetilde D_{n+1}^{\pi})}{X_{n+1}^{\pi}}\Big)^2\Big]
	-\frac{\gamma_n(d)}{2}\Big({\bbe_{n,\tilde d}\Big[\frac{g_{n+1}(\widetilde D_{n+1}^{\pi})}{x}\Big]}\Big)^2{\Big)}\Big\},
	\end{align} 
	for $0\leq n<T$ and $\tilde d=(x,d)\in\widetilde{\mathcal{D}}_n$, with terminal condition 
	\begin{align*} 
	V_{T}(\tilde d)&=0, \quad \tilde d\in\widetilde{\mathcal{D}}_T.
	\end{align*}
	Herein, for any fixed $k=0,1,\dots,T$ and $\tilde d'=(x',d')\in\widetilde{\mathcal{D}}_k$, the function sequence $(f_{n,k}(\cdot;\tilde d'))_{0\leq n\leq T}$, where $f_{n,k}(\cdot;\tilde d'):\widetilde{\mathcal{D}}_n\mapsto\bbr$, is determined by the recursion 
	\begin{alignat*}{3}
	f_{n,k}(\tilde d;\tilde d') &=\bbe_{n,\tilde d}[f_{n+1,k}(\widetilde D_{n+1}^{\pi^*};\tilde d')],\quad &&\tilde d\in\widetilde{\mathcal{D}}_n, \quad 0\leq n<T,\\
	f_{T,k}(\tilde d;\tilde d') &=  \frac{{x}}{x'}-1-\frac{\gamma_k(d')}{2} \Big(\frac{{x}}{x'}\Big)^2, \quad &&\tilde d\in\widetilde {\mathcal{D}}_T,
	\end{alignat*}
	and the function sequence $(g_n)_{0\leq n\leq T}$, where $g_n:\widetilde{\mathcal{D}}_n\mapsto\bbr$, is determined by the recursion
	\begin{align*}
	g_{n}(\tilde d)&=\bbe_{n,\tilde d}[g_{n+1}(\widetilde D_{n+1}^{\pi^*})],\quad \tilde d\in\widetilde{\mathcal{D}}_n, \quad 0\leq n<T,\\
	g_T(\tilde d) &= x,\quad \tilde d\in\widetilde{\mathcal{D}}_T.
	\end{align*}
	Furthermore, the function sequences admit the probabilistic {representations}  
	\begin{align*}
	f_{n,k}(\tilde d;\tilde d') &= \bbe_{n,\tilde d}\Big[\frac{X_T^{\pi^*}}{x'}-1-\frac{\gamma_k(d')}{2}\Big(\frac{X_T^{\pi^*}}{x'}\Big)^2\Big],
	\qquad
	g_n(\tilde d) = \bbe_{n,d}[X_T^{\pi^*}].
	\end{align*}
\end{prop}

\paragraph{Proof of Proposition \ref{PropHJB}:}

{
	
	We write the objective functional at time $n$ as
	\begin{align*}
	J_{n}(\tilde d;\pi) &=\bbe_{n,\tilde d}[F_n(\tilde d,X_T^{\pi})]+G_n(\tilde d,\bbe_{n,\tilde d}[X_T^{\pi}]),
	\end{align*}
	where, for $\tilde d=(x,d)\in\widetilde{\mathcal{D}}_n$ and $y\in\bbr$, 
	\begin{align}\label{FG}
	F_n(\tilde d,y) &:= \frac{y}{x}-1-\frac{\gamma_n(\tilde d)}{2} \Big(\frac{y}{x}\Big)^2,\qquad 
	G_n(\tilde d,y) := \frac{\gamma_n(\tilde d)}{2} \Big(\frac{y}{x}\Big)^2.
	\end{align}
	The proof now consists of two parts. First, we derive the recursive equation satisfied by the sequence of objective functionals for any admissible control law $\pi$. Then, we derive the system of equations that an optimal control law $\pi^*$ must satisfy. 
	
	\noindent \emph{Step 1: {Recursion for $J_{n}(\tilde d;\pi)$.} } For a given admissible control law $\pi$, and fixed $0\leq k\leq T$ and $\tilde d'\in \widetilde{\mathcal{D}}_k$, we define the function sequences
	\begin{equation*}
	\begin{alignedat}{3}\label{fg1}
	f_{n,k}^{\pi}(\tilde d;\tilde d') &:= \bbe_{n,\tilde d}[F_k(\tilde d',X_T^{\pi})],\quad &&\tilde d\in\widetilde{\mathcal{D}}_n,\quad 0\leq n\leq T,\\
	\qquad g_n^{\pi}(\tilde d) &:= \bbe_{n,\tilde d}[X_T^{\pi}],\quad &&\tilde d\in\widetilde{\mathcal{D}}_n,\quad 0\leq n\leq T,
	\end{alignedat}
	\end{equation*}
	and note that, by the law of iterated expectations,
	\begin{align}\label{MGnew}
	\begin{split}
	f_{n,k}^{\pi}(\tilde d;\tilde d')  
	&=\bbe_{n,\tilde d}[f_{n+1,k}^{\pi}(\widetilde D_{n+1}^{\pi};\tilde d')],\qquad 
	g_n^{\pi}(\tilde d) 
	=\bbe_{n,\tilde d}[g_{n+1}^{\pi}(\widetilde D_{n+1}^{\pi})].
	\end{split}
	\end{align}
	For $\widetilde D_{n+1}\in\widetilde{\mathcal{D}}_{n+1}$, the objective functional at time $n+1$ can then be written as
	\begin{align*}
	J_{n+1}(\widetilde D_{n+1};\pi)&=\bbe_{n+1,\widetilde D_{n+1}}[F_{n+1}(\widetilde D_{n+1},X_T^{\pi})]+G_{n+1}(\widetilde D_{n+1},\bbe_{n+1,\widetilde D_{n+1}}[X_T^{\pi}])\\
	&= f_{n+1,n+1}^{\pi}(\widetilde D_{n+1};\widetilde D_{n+1}) + G_{n+1}(\widetilde D_{n+1},g_{n+1}^{\pi}(\widetilde D_{n+1})).
	\end{align*}
	{Taking expectations} with respect to $\bbp_{n,\tilde d}$ and applying the control law $\pi$ at time $n$ yields 
	\begin{align*}
	\bbe_{n,\tilde d}[J_{n+1}(\widetilde D_{n+1}^{\pi};\pi)] &= \bbe_{n,\tilde d}[f_{n+1,n+1}^{\pi}(\widetilde D_{n+1}^{\pi};\widetilde D_{n+1}^{\pi})]
	+ \bbe_{n,\tilde d}[G_{n+1}(\widetilde D_{n+1}^{\pi},g_{n+1}^{\pi}(\widetilde D_{n+1}^{\pi})].
	\end{align*}
	Adding and subtracting $J_{n}(\tilde d;\pi)$ gives
	\begin{align*}
	\bbe_{n,\tilde d}[J_{n+1}(\widetilde D_{n+1}^{\pi};\pi)]
	= J_{n}(\tilde d;\pi)
	&+\bbe_{n,\tilde d}[f_{n+1,n+1}^{\pi}(\widetilde D_{n+1}^{\pi};\widetilde D_{n+1}^{\pi})]-\bbe_{n,\tilde d}[F_n(\tilde d,X_T^{\pi})] \\
	&+\bbe_{n,\tilde d}[G_{n+1}(\widetilde D_{n+1}^{\pi},g_{n+1}^{\pi}(\widetilde D_{n+1}^{\pi}))]
	-G_{n}(\tilde d,\bbe_{n,\tilde d}[X_T^{\pi}]).
	\end{align*}
	{Using} (\ref{fg1})-(\ref{MGnew}), we then obtain 
	\begin{align}\label{Jrec}
	\begin{split}
	J_{n}(\tilde d;\pi)
	= \bbe_{n,\tilde d}[J_{n+1}(\widetilde D_{n+1}^{\pi};\pi)]
	&-\Big(\bbe_{n,\tilde d}[f_{n+1,n+1}^{\pi}(\widetilde D_{n+1}^{\pi};\widetilde D_{n+1}^{\pi})] -\bbe_{n,\tilde d}[f_{n+1,n}^{\pi}(\widetilde D_{n+1}^{\pi};\tilde d)]\Big)\\
	&-\Big(\bbe_{n,\tilde d}[G_{n+1}(\widetilde D_{n+1}^{\pi},g_{n+1}^{\pi}(\widetilde D_{n+1}^{\pi})] -G_{n}(\tilde d,\bbe_{n,\tilde d}[g_{n+1}^{\pi}(\widetilde D_{n+1}^{\pi})])\Big).
	\end{split}
	\end{align}
	
	\noindent\emph{Step 2: Recursion for $V_{n}(\tilde d)$.} Assume the existence of an optimal control law $\pi^*$ and consider a control law $\pi$ that coincides with $\pi^*$ after time $n$, {i.e.,} $\pi_k\equiv\pi^*_k$, for $k=n+1,\dots,T-1$.
	By definition, we then have
	\begin{align*}
	J_{n}(\tilde d;\pi)&\leq V_{n}(\tilde d) = J_{n}(\tilde d;\pi^*), \quad \tilde d\in\widetilde{\mathcal{D}}_n.
	\end{align*}
	Recall the definition of function sequences in (\ref{fg1}). For the optimal control law $\pi^*$, let 
	\begin{align}\label{fg2}
	\begin{split}
	f_{n,k}(\tilde d;\tilde d') &:= f_{n,k}^{\pi^*}(\tilde d;\tilde d'),\qquad 
	g_n(\tilde d) := g_n^{\pi^*}(\tilde d). 
	\end{split}
	\end{align}
	{Since} $\pi$ and $\pi^*$ coincide after time $n$,  for any $\widetilde D_{n+1}\in\widetilde{\mathcal{D}}_{n+1}$ {we have that} 
	\begin{align*}
	J_{n+1}(\widetilde D_{n+1};\pi) &= V_{n+1}(\widetilde D_{n+1}),\quad 
	f_{n+1,k}(\widetilde D_{n+1};\tilde d') = f_{n+1,k}^{\pi}(\widetilde D_{n+1};\tilde d'),\quad 
	g_{n+1}(\widetilde D_{n+1}) = g_{n+1}^{\pi}(\widetilde D_{n+1}).
	\end{align*}
	In turn, using the recursion (\ref{Jrec}) for $J_{n}(d;\pi)$, we may write for $\tilde d\in\widetilde{\mathcal{D}}_n$,
	\begin{align*}
	V_{n}(\tilde d)
	=\sup_{\pi}\Big\{ 
	\bbe_{n,\tilde d}[V_{n+1}(\widetilde D_{n+1}^{\pi})]
	&-\Big(\bbe_{n,\tilde d}[f_{n+1,n+1}(\widetilde D_{n+1}^{\pi};\widetilde D_{n+1}^{\pi})] -\bbe_{n,\tilde d}[f_{n+1,n}(\widetilde D_{n+1}^{\pi};\tilde d)]\Big)\nonumber\\
	& -\Big(\bbe_{n,\tilde d}[G_{n+1}(\widetilde D_{n+1}^{\pi},g_{n+1}(\widetilde D_{n+1}^{\pi})] -G_n(\tilde d,\bbe_{n,\tilde d}[g_{n+1}(\widetilde D_{n+1}^{\pi})])\Big)\Big\},\nonumber
	\end{align*}
	with the terminal condition $V_{T}(\tilde d)=0$, for $\tilde d\in\mathcal{D}_T$.
	The recursions and probabilistic {representations of} $(f_{n,k}(\cdot;\tilde d'))_{0\leq n\leq T}$ and $(g_{n})_{0\leq n\leq T}$ then follow from (\ref{FG})-(\ref{MGnew}), and (\ref{fg2}). 
	\hfill\qed
	
	\paragraph{Proof of Proposition \ref{prop1}:} 
	Assuming the existence of an optimal control law $\pi^*$, the value function at time $n+1$ satisfies 
	\begin{align*}
	V_{n+1}(\widetilde D_{n+1}) = f_{n+1,n+1}(\widetilde D_{n+1};\widetilde D_{n+1})  + \frac{\gamma_{n+1}( D_{n+1})}{2} \left(\frac{g_{n+1}(\widetilde D_{n+1})}{X_{n+1}}\right)^2.
	\end{align*}
	{Plugging the above expression} into the HJB equation (\ref{HJB}) gives 
	\begin{align}\label{HJBnew}
	\begin{split}
	V_{n}(\tilde d) =\sup_{\pi}\Big\{ \bbe_{n,\tilde d}[f_{n+1,n}(\widetilde D_{n+1}^{\pi};\tilde d)]
	+\frac{\gamma_n(d)}{2}\left(\frac{\bbe_{n,\tilde d}[g_{n+1}(\widetilde D_{n+1}^{\pi})]}{x}\right)^2\Big\}.
	\end{split}
	\end{align}
	Next, we look for a candidate optimal control law of form (\ref{pi_x}), i.e., $\pi_n(\tilde d)= \tilde\pi_{n}(d) x$. 
	{It then follows from Lemma} \ref{lem_ab}, {along with} (\ref{FG}), (\ref{fg1}), and (\ref{fg2}) that, for this candidate optimal law, 
	\begin{align*}
	f_{n+1,n}(\tilde d;\tilde d') &= a_{n+1}^{\pi}(d) \frac{x}{x'}-1-\frac{\gamma_{n}(\tilde d')}{2} b_{n+1}^{\pi}(d) \Big(\frac{x}{x'}\Big)^2,\\
	g_{n+1}(\tilde d) &= a_{n+1}^{\pi}(d) x.
	\end{align*}
	Plugging {the above expression} into (\ref{HJBnew}), using the wealth dynamics (\ref{dX}) and, for brevity, writing $R$ for $R_{n+1}$, $\tilde\pi_n$ for $\tilde\pi_n(d)$, $a_{n+1}^{\pi}$ for $a_{n+1}^{\pi}(D_{n+1})$, and $b_{n+1}^{\pi}$ for $b_{n+1}^{\pi}(D_{n+1})$, yields
	\begin{align}\label{hDef}
	V_{n}(\tilde d) 
	&= \sup_{\pi}\Big\{ \bbe_{n,\tilde d}\Big[a^{\pi}_{n+1} \frac{X_{n+1}^{\pi}}{x}-1-\frac{\gamma_n( d)}{2} b^{\pi}_{n+1} \Big(\frac{X_{n+1}^{\pi}}{x}\Big)^2\Big] 
	+ \frac{\gamma_n( d)}{2}\Big({\bbe_{n,\tilde d}\Big[a^{\pi}_{n+1} \frac{X_{n+1}^{\pi}}{x}\Big]}\Big)^2\Big\}\\
	&= \sup_{\pi}\Big\{\bbe_{n,\tilde d}\Big[a^{\pi}_{n+1}(R+\widetilde Z_{n+1}\tilde\pi_n)-1-\frac{{\gamma_n( d)}}{2} b^{\pi}_{n+1}(R+\widetilde Z_{n+1}\tilde\pi_n)^2\Big]
	+ \frac{\gamma_n( d)}{2}\Big(\bbe_{n,\tilde d}[a^{\pi}_{n+1}(R+\widetilde Z_{n+1}\tilde\pi_n)]\Big)^2\Big\}\\
	&= \sup_{\pi}\Big\{\bbe_{n,\tilde d}\Big[a^{\pi}_{n+1}(R+\widetilde Z_{n+1}\tilde\pi_n)-1-\frac{{\gamma_n( d)}}{2} b^{\pi}_{n+1}(R^2+2R\tilde\pi_n\widetilde Z_{n+1}+\tilde\pi_n^2\widetilde Z_{n+1}^2)\Big]\\
	&\quad\quad + \frac{\gamma_n( d)}{2}\Big(R^2(\bbe_{n,\tilde d}[a^{\pi}_{n+1}])^2 + 2R\tilde\pi_n\bbe_{n,\tilde d}[a^{\pi}_{n+1}]\bbe_{n,\tilde d}[a^{\pi}_{n+1}\widetilde Z_{n+1}] + \tilde\pi_n^2(\bbe_{n,\tilde d}[a^{\pi}_{n+1}\widetilde Z_{n+1}])^2\Big)\Big\} 
	{=: \sup_{\pi}h_n(\tilde\pi_n),}
	\end{align}
	where $h_n$ is a second order polynomial with derivative
	\begin{align*}
	h_n'(\tilde\pi_n)  
	&= \bbe_{n,\tilde d}[a^{\pi}_{n+1}\widetilde Z_{n+1}] -R\gamma_n(d)\big(\bbe_{n,\tilde d}[b^{\pi}_{n+1}\widetilde Z_{n+1}] - \bbe_{n,\tilde d}[a^{\pi}_{n+1}]\bbe_{n,\tilde d}[a^{\pi}_{n+1}\widetilde Z_{n+1}]\big)\\
	&\quad - \gamma_n(d)\big( \bbe_{n,\tilde d}[b^{\pi}_{n+1}\widetilde Z_{n+1}^2] - (\bbe_{n,\tilde d}[a^{\pi}_{n+1}\widetilde Z_{n+1}])^2\big)\tilde\pi_n. 
	\end{align*}
	By Jensen's inequality, $\bbe_{n,\tilde d}[b^{\pi}_{n+1}\widetilde Z_{n+1}^2] - (\bbe_{n,\tilde d}[a^{\pi}_{n+1}\widetilde Z_{n+1}])^2 > 0$, and solving for $h_n'(\tilde\pi_n)=0$ yields the formulation in Proposition \ref{prop1}.   
	One can then easily check that this solves the HJB equation (\ref{HJB}). To obtain the allocation formula (\ref{pi_hat}), we first use (\ref{abRecursions}) and the law of iterated expectations to rewrite the identities in (\ref{muamub}),
	\begin{align*}
	\mu_n^a(d) &= \bbe_{n,d}[\bbe_{n+1,D_{n+1}}[1+r_{n+1,T}^{\pi^*}]] = \bbe_{n,d}[1+r_{n+1,T}^{\pi^*}], \\
	\mu_n^{az}(d) &= \bbe_{n,d}[\widetilde Z_{n+1}\bbe_{n+1,D_{n+1}}[1+r_{n+1,T}^{\pi^*}]] = \bbe_{n,d}[\widetilde Z_{n+1}(1+r_{n+1,T}^{\pi^*})], \\
	\mu_n^{bz}(d) &= \bbe_{n,d}[\widetilde Z_{n+1}\bbe_{n+1,D_{n+1}}[(1+r_{n+1,T}^{\pi^*})^2]] = \bbe_{n,d}[\widetilde Z_{n+1}(1+r_{n+1,T}^{\pi^*})^2], \\
	\mu_n^{bz^2}(d) &= \bbe_{n,d}[\widetilde Z_{n+1}^2\bbe_{n+1,D_{n+1}}[(1+r_{n+1,T}^{\pi^*})^2]] = \bbe_{n,d}[\widetilde Z_{n+1}^2(1+r_{n+1,T}^{\pi^*})^2].
	\end{align*}
	It then follows from the definition of variance and covariance that 
	\begin{align*}
	\mu_n^{bz^2}(d)- (\mu_n^{az}(d))^2 &= Var_{n,d}[\widetilde Z_{n+1}(1+r_{n+1,T}^{\pi^*})], \\
	\mu_n^{bz}(d) - \mu_n^a(d)\mu_n^{az}(d) &= Cov_{n,d}[\widetilde Z_{n+1}(1+r_{n+1,T}^{\pi^*}),1+r_{n+1,T}^{\pi^*}],
	\end{align*}
	and (\ref{pi_hat}) is obtained by plugging the above into formula (\ref{uhatprop2}). \hfill\qed
	
}

\paragraph{Proof of Proposition \ref{PropUhat}:} It follows directly from the proof of Proposition \ref{prop1}. \hfill\qed 

\paragraph{Proof of Lemma \ref{lem0}:} Observe that 
no short-selling and borrowing implies $0\leq \tilde\pi_n^*(d) \leq 1$. Let
\begin{align*}
A_n(d):=\{1+\widetilde Z_{n+1}\tilde\pi_n^*(d)<0\}, 
\end{align*}
which, for any positive wealth level at time $n$, is the event of the wealth becoming negative at time $n+1$. Recall from (\ref{abr}) the return moments $a_n(d)$ and $b_n(d)$ for the optimal strategy $\pi^*$, and, for $x>0$, define for $\pi^o$ the return moments
\begin{align*}
  a_n^o(d) := \bbe_{n,x,d}[1+r_{n,T}^{\pi^o}], 
  \qquad
 b_n^o(d) := \bbe_{n,x,d}\big[(1+r_{n,T}^{\pi^o})^2\big], 
\end{align*}
which are independent of the value of $x$. 
This is because the probability of the wealth becoming negative, $\bbp_{n,x,d}\big(X_k^{\pi^o}<0 \text{ for some } k\in\{n+1,\dots,T-1\}\big)$, is independent of the value of $x>0$.

We use induction to show that, for 
any $d\in\mathcal{D}_n$, 
\begin{align*}
| a^o_{n}(d) - a_{n}(d)| \leq 2\big(T-(n+1)\big)K^{T-n}\bbp_{n,d}\big(\widetilde Z_{n+1}<-1\big) := K_n.
\end{align*} 
where $K := 
\max_{y\in\mathcal{Y}}\big(1+r(y)+\bbe\big[\big|\widetilde Z_1\big||Y_0=y\big]\big)$ is a constant. 
First, for $n=T-1$, we have $a_n^o(d) = a_n(d)$. 
Then, for a general $n\in\{0,1,\dots,T-2\}$, we have
\begin{align*}
a^o_n(d) 
&= \bbe_{n,d}\big[(R_{n+1}+\widetilde Z_{n+1}\tilde\pi_n^*(d)) a^o_{n+1}(D_{n+1}){\bf 1}_{A_n^c}\big] + \bbe_{n,d}\big[(R_{n+1}+\widetilde Z_{n+1}\tilde\pi_n^*(d)) a^o_{n+1}(D_{n+1}){\bf 1}_{A_n}\big] \\
&= \bbe_{n,d}\big[(R_{n+1}+\widetilde Z_{n+1}\tilde\pi_n^*(d)) a^o_{n+1}(D_{n+1})\big]
- \bbe_{n,d}\big[(R_{n+1}+\widetilde Z_{n+1}\tilde\pi_n^*(d)) a^o_{n+1}(D_{n+1}){\bf 1}_{A_n}\big] \\
&\quad+ \bbe_{n,d}\big[(R_{n+1}+\widetilde Z_{n+1}\tilde\pi_n^*(d))\Big(\prod_{k=n+1}^{T-1}R_{k+1}\Big){\bf 1}_{A_n}\big],
\end{align*}
so, assuming the conjecture to hold for $n+1$, and by using $0\leq\tilde\pi_n^*\leq 1$, $R_{n+1}=1+r_{n+1}$, and that $\widetilde Z_{n+1}$ has mean $\mu_{n+1}-r_{n+1}$, we obtain
\begin{align*}
| a^o_n(d)-a_n(d)| &\leq K_{n+1}\bbe_{n,d}\big[|R_{n+1}+\widetilde Z_{n+1}\tilde\pi_n^*(d)|\big]
+ \bbe_{n,d}\big[|R_{n+1}+\widetilde Z_{n+1}\tilde\pi_n^*(d)| a^o_{n+1}(D_{n+1}){\bf 1}_{A_n}\big]\\
&\quad + \bbe_{n,d}\Big[|R_{n+1}+\widetilde Z_{n+1}\tilde\pi_n^*(d)|\Big(\prod_{k=n+1}^{T-1}R_{k+1}\Big){\bf 1}_{A_n}\Big] \\
&\leq K_{n+1}K + 2K^{T-n}\bbp_{n,d}(\widetilde Z_{n+1}<-1) \\ 
&\leq 2(T-(n+2))K^{T-n}\bbp_{n,d}(\widetilde Z_{n+1}<-1) + 2K^{T-n}\bbp_{n,d}(\widetilde Z_{n+1}<-1) \\
&= K_n. 
\end{align*}
A similar result can be derived for the second moment $b^o(d)$. The results for the variance and objective functional then follow. \hfill\qed

\paragraph{Proof of Lemma \ref{LemMu}:} The formulas for $\mu_n^{az}(d)$ follow directly from (\ref{muamub})-(\ref{abRecursions}) and the definition of the random variable $D_{n+1}$. In part (a), the value of $\tau_{n+1}$ is determined by $(Y_{(n+1)},Z_{(n+1)},\epsilon_{(n+1)})$, because the interaction schedule $(T_k)_{k\geq 0}$ is adapted to the filtration $(\calF_n)_{n\geq 0}$ in (\ref{calF}). Conditionally on $\tau_{n+1}=n+1$, the value of $\xi_{n+1}$ is determined by $(Y_{(n+1)},Z_{(n+1)},\epsilon_{(n+1)})$ for the same reason. In part (b), (\ref{gammaCex}) and (\ref{gammaCZ}) yield that if $\tau_{n+1}=n+1$, then
\begin{align*}
\xi_{n+1} 
= e^{\eta_{n+1}-\eta_{\tau_n}}\xi_{\tau_n} e^{\sum_{k=\tau_n}^{n}\epsilon_{k+1}} \frac{\bar\gamma_{n+1}(Y_{n+1})}{\bar\gamma_{\tau_n}(Y_{\tau_n})}\frac{\gamma_{n+1}^{Z}}{\gamma_{\tau_n}^{Z}},
\end{align*}
where $\gamma_{n+1}^{Z}$ and $\gamma_{\tau_n}^{Z}$ depend on $\{Y_{\tau_n-\phi},\dots,Y_n\}$ and $\{Z_{\tau_n-\phi+1},\dots,Z_{n+1}\}$, and $\tau_n=\phi-1$ is the interaction time prior to $n+1$. 
\hfill\qed

\paragraph{Proof of (\ref{u_approx}):} The optimal allocation in Proposition \ref{prop1} can be written as
\begin{align*}
\tilde\pi^*_n(d) 
&= \frac{\tilde\mu_{n+1}}{\gamma_n\sigma^2_{n+1}}\frac{\bbe_{n,d}\Big[\frac{\widetilde Z_{n+1}}{\tilde\mu_{n+1}}\tilde r_{n+1,T}^{\pi^*}\Big]-R_{n+1}\gamma_n\Big(\bbe_{n,d}\Big[\frac{\widetilde Z_{n+1}}{\tilde\mu_{n+1}}(\tilde r_{n+1,T}^{\pi^*})^2 \Big]-\bbe_{n,d}\Big[\frac{\widetilde Z_{n+1}}{\tilde\mu_{n+1}}\tilde r_{n+1,T}^{\pi^*}\Big]\bbe_{n,d}\big[\tilde r_{n+1,T}^{\pi^*}\big]\Big)}{\bbe_{n,d}\Big[\Big(\frac{\widetilde Z_{n+1}}{\sigma_{n+1}}\tilde r_{n+1,T}^{\pi^*}\Big)^2\Big]
	-\Big(\bbe_{n,d}\Big[\frac{\widetilde Z_{n+1}}{\sigma_{n+1}}\tilde r_{n+1,T}^{\pi^*}\Big]\Big)^2}.
\end{align*}
This simplifies to formula (\ref{u_approx}) under the assumption of Corollary \ref{cor1},  
in which case the expected values in the previous formula factor into products of expectations. \hfill\qed
\section{Proofs and Auxiliary Results for Section \ref{SectionRegret}}\label{AppendixC}

\paragraph{Proof of Proposition \ref{propRegret}:} 

From the definitions of $\gamma_n$ and $\gamma_n^C$, it follows that
\begin{align}\label{R2}
\mathcal{R}(\phi,\beta)
&=\frac{1}{T}\sum_{n=0}^{T-1}\bbe\Big[\Big|\frac{\gamma_n^{id}}{\gamma_{\tau_n}^{id}\gamma_{\tau_n}^Z}-1\Big|\Big]. 
\end{align}
Let $Y_0=y_0$ and extend the probability space $(\Omega,\calF,\bbp)$ to include an independent sequence of random variables $(Z_{n})_{n\leq 0}$, such that $Z_n\sim\mathcal{N}(\tilde\mu(y_0),\sigma^2(y_0))$. For a given $n$, define the event $A_n:=\{Y_{\tau_n-\phi}=\dots=Y_{\tau_n-1}=y_0\}$, on which the economic state variable is fixed. We then have
\begin{align*}
\bbe\Big[\Big|\frac{\gamma_n^{id}}{\gamma_{\tau_n}^{id}\gamma_{\tau_n}^Z}-1\Big|\Big]
&= \bbe\Big[\Big|e^{\beta\big(\frac{1}{\phi}\sum_{k=\tau_n-\phi}^{\tau_n-1}  Z_{k+1}-\mu_{k+1}\big)+\sum_{k=\tau_n}^{n-1}\epsilon_{k+1}}-1\Big|A_n\Big]
+  O\big(1-P_{y_0,y_0}^{\tau_n-1}\big),
\end{align*}
where $P_{y_0,y_0}^{\tau_n-1}:=(P^{\tau_n-1})_{y_0,y_0}$ is the probability of staying in state $y_0$ from time zero to time $\tau_n-1$. 
For the first term appearing in the exponent, we have
\begin{align*}
\Big.\beta\Big(\frac{1}{\phi}\sum_{k=\tau_n-\phi}^{\tau_n-1}  Z_{k+1}-\mu_{k+1}\Big) \Big| A_n
\sim\mathcal{N}\Big(0,\frac{\beta^2\sigma_0^2}{\phi}\Big).
\end{align*}
For the second term, we define the random variable $J:=|\{\tau_n\leq k\leq n-1:\epsilon_{k+1}\neq 0\}| \sim \text{Bin}(n-\tau_n,p_{\epsilon})$, 
which is equal to the number of idiosyncratic risk aversion jumps between times $\tau_n$ and $n$. Hence,
\begin{align*}
\sum_{k=\tau_n}^{n-1}\epsilon_{k+1}\Big|\{J=j\} \sim\mathcal{N}\Big(-j\frac{\sigma_{\epsilon}^2}{2},j\sigma_{\epsilon}^2\Big).
\end{align*}
For $Z\sim\mathcal{N}(0,1)$, $a\in\bbr$, and $b\geq 0$, we have $\bbe[|e^{a+b Z}-1|] = b\bbe[|Z|] + O(|a| + b^2)$. Therefore,  conditioning on the value of $J$ and using the absolute Gaussian moment formula gives
\begin{align*}
\bbe\Big[\Big|\frac{\gamma_n^{id}}{\gamma_{\tau_n}^{id}\gamma_{\tau_n}^Z}-1\Big|\Big]
= \sqrt{\frac{2}{\pi}}\Big((1-p_{\epsilon})^{n-\tau_n}\frac{\beta\sigma_0}{\sqrt{\phi}}
+ (n-\tau_n)p_{\epsilon}(1-p_{\epsilon})^{n-\tau_n-1}\sqrt{\frac{\beta^2\sigma_0^2}{\phi}+\sigma_{\epsilon}^2}\Big) + \Theta + O(1-P_{y_0,y_0}^{\tau_n-1}), 
\end{align*}
{where we have denoted by $\Theta$ a term of order $O(p^2\sigma_{\epsilon}^2 + \beta^2)$. Next, plugging the above expression into} (\ref{R2}), {and noting that the sum is the same for each interval bounded by consecutive times of interaction, we obtain} 
\begin{align*}
\mathcal{R}(\phi,\beta)
&\geq\frac{T_{\phi}}{T}\sqrt{\frac{2}{\pi}}\frac{1}{\phi}\Big(\frac{\beta\sigma_0}{\sqrt{\phi}}\sum_{n=0}^{\phi-1}(1-p_{\epsilon})^{n}
+ \sqrt{\frac{\beta^2\sigma_0^2}{\phi}+\sigma_{\epsilon}^2}\sum_{n=0}^{\phi-1}np_{\epsilon}(1-p_{\epsilon})^{n-1}\Big) + \Theta +O(1-P_{y_0,y_0}^{T-\phi-1})\\
&=\frac{T_{\phi}}{T}\sqrt{\frac{2}{\pi}}\frac{1}{\phi}\Big(\frac{\beta\sigma_0}{\sqrt{\phi}}\frac{1-(1-p_{\epsilon})^{\phi}}{p_{\epsilon}}
+ \sqrt{\frac{\beta^2\sigma_0^2}{\phi}+\sigma_{\epsilon}^2} \frac{1-(1-p_{\epsilon})^{\phi}-\phi(1-p_{\epsilon})^{\phi-1}p_{\epsilon}}{p_{\epsilon}}\Big) + \Theta +O(1-P_{y_0,y_0}^{T-\phi-1}) \\
&= \frac{T_{\phi}}{T}\widetilde{\mathcal{R}}(\phi,\beta) +\Theta +O(1-P_{y_0,y_0}^{T-1}),
\end{align*}
where the final equality follows from using a third-order Taylor approximation for $(1-p_{\epsilon})^{\phi}$ and the fact that $P_{y_0,y_0}^{T-\phi-1}$ is increasing in $\phi\geq 1$. {The upper bound in} (\ref{genT}) {follows from a symmetric argument.} 

Next, {to show that there exists a unique minimizing $\phi$ for $\widetilde{\mathcal{R}}(\phi,\beta)$,}  
	we define the function
\begin{align*} 
f(\phi) &:= \sqrt{\frac{2}{\pi}}\widetilde{\mathcal{R}}(\phi,\beta) = \frac{\beta\sigma_0}{\sqrt{\phi}}\Big(1  - \frac{\phi-1}{2}p_{\epsilon}\Big)  + \sqrt{\frac{\beta^2\sigma_0^2}{\phi}+\sigma_{\epsilon}^2}\frac{\phi-1}{2}p_{\epsilon}.
\end{align*}
Observe that $f\geq 0$ and that if $p_{\epsilon}=0$ and $\beta=0$, then $f\equiv 0$. If $p_{\epsilon}=0$ and $\beta>0$, then $f$ is strictly decreasing for $\phi\geq 1$, and converges to zero as $\phi\to\infty$. If $\beta=0$ and $p_{\epsilon}>0$, then $f$ is strictly increasing for $\phi\geq 1$, and equal to zero for $\phi=1$. Furthermore,
\begin{align*} 
f'(\phi) 
&= -\half\frac{\beta\sigma_0}{\phi^{3/2}}\Big(1  - \frac{\phi-1}{2}p_{\epsilon}\Big) - \frac{\beta\sigma_0}{\sqrt{\phi}}\frac{p_{\epsilon}}{2} 
- \frac{1}{2\sqrt{\frac{\beta^2\sigma_0^2}{\phi}+\sigma_{\epsilon}^2}}\frac{\beta^2\sigma_0^2}{\phi^2}\frac{\phi-1}{2}p_{\epsilon} + \sqrt{\frac{\beta^2\sigma_0^2}{\phi}+\sigma_{\epsilon}^2}\frac{p_{\epsilon}}{2} \\
&= \half\frac{\beta\sigma_0 p_{\epsilon}}{\phi^2}\Big(-\frac{\sqrt{\phi}}{p_{\epsilon}} + \half\Big(1
- \frac{1}{\sqrt{1+\frac{\sigma_{\epsilon}^2\phi}{\beta^2\sigma_0^2}}}\Big)\sqrt{\phi}({\phi-1}) + \Big(\sqrt{1+\frac{\sigma_{\epsilon}^2\phi}{\beta^2\sigma_0^2}}- 1\Big)\phi^{3/2}\Big) 
=: \half\frac{\beta\sigma_0 p}{\phi^2}g(\phi),
\end{align*}
{and it follows that the derivative of $f$ at $\phi=1$ satisfies}
\begin{align*}
f'(1) = \half{\beta\sigma_0 p_{\epsilon}}\Big( \sqrt{1+\frac{\sigma_{\epsilon}^2}{\beta^2\sigma_0^2}}- 1-\frac{1}{p_{\epsilon}}\Big) \geq 0
&\quad\Longleftrightarrow\quad \frac{p_{\epsilon}\sigma_{\epsilon}^2}{\beta^2\sigma_0^2} \geq  2 + \frac{1}{p_{\epsilon}}.
\end{align*} 
Observe that, since the value $p_{\epsilon}$ is in general small, we have
\begin{align}\label{phiCond}
\frac{p_{\epsilon}\sigma_{\epsilon}^2}{\beta^2\sigma_0^2} \geq  2 + \frac{1}{p_{\epsilon}}
\quad\Longleftrightarrow\quad 
\frac{p_{\epsilon}\sigma_{\epsilon}}{\beta\sigma_0} \geq  \sqrt{1+2p_{\epsilon}} \approx 1,
\end{align} 
which shows that $p_{\epsilon}\sigma_{\epsilon}/({\beta\sigma_0})> 1$ is, at this approximation level, a sufficient and necessary condition for $f'(1)<0$.
Furthermore, 
\begin{align*}
g'(\phi) & =-\half\frac{\phi^{-1/2}}{p_{\epsilon}} + \Big[\half\Big(1
- \frac{1}{\sqrt{1+\frac{\sigma_{\epsilon}^2\phi}{\beta^2\sigma_0^2}}}\Big)\sqrt{\phi}({\phi-1})\Big]'
+ \frac{3}{2}\Big(\sqrt{1+\frac{\sigma_{\epsilon}^2\phi}{\beta^2\sigma_0^2}}- 1\Big)\phi^{1/2}
+ \frac{1}{2\sqrt{1+\frac{\sigma_{\epsilon}^2\phi}{\beta^2\sigma_0^2}}} \frac{\sigma_{\epsilon}^2}{\beta^2\sigma_0^2}\phi^{3/2} \\
& \geq \half\phi^{-1/2}\Big(-\frac{1}{p_{\epsilon}} 
+ \Big(\sqrt{1+\frac{\sigma_{\epsilon}^2}{\beta^2\sigma_0^2}}- 1\Big)\phi\Big) 
\geq \half\phi^{-1/2}g(1).
\end{align*}
It then follows that if $g(1)\geq 0$, then $g'(\phi)\geq 0$, for all $\phi\geq 1$. {Hence, } $f'(\phi)\geq 0$ for all $\phi\geq 1$, and $f$ {is} minimized for $\phi=1$. If $g(1)<0$, {and thus} $f'(1)<0$, it can be shown that $g$ is strictly convex and $g(\phi)\to\infty$, as $\phi\to\infty$. Hence, there exists a unique point where $g(\phi)=0$, and where $f$ is minimized. Finally, from the form of the function $f$, it is easy to see the sign of the derivatives of the optimal value of $\phi$ with respect to $\beta, \sigma_0, p_{\epsilon}$, and $\sigma_{\epsilon}$. \hfill\qed

\paragraph{Proof of (\ref{SRapprox}):} For the two investment strategies we denote the coefficient $\mu_n^a(D_n)$ in (\ref{muamub}) by $\mu_n^a(\gamma_n)$ and $\mu_n^a(\gamma_n^C)$, and show inductively that 
\begin{align*} 
\bbe[|\mu_{n}^a(\gamma_n) - \mu_n^a(\gamma_n^C)|] &= O((\phi-1)p_{\epsilon}\sigma_{\epsilon}^2) + O\Big(\frac{\beta^2\sigma^2}{\phi^2}\Big).
\end{align*}
Analogous identities can be obtained for $\mu_{n}^{b}$, $\mu_{n}^{az}$, $\mu_{n}^{az^2}$, and $\mu_{n}^{bz}$. One may, then, inductively show that
\begin{align*}
\tilde\pi^*_n(\gamma_n) 
&= \frac{\gamma_n^C}{\gamma_n}\tilde\pi^*_n(\gamma_n^C) +O((\phi-1)p_{\epsilon}\sigma_{\epsilon}^2) + O\Big(\frac{\beta^2\sigma^2}{\phi^2}\Big),
\end{align*}
and the result follows.\hfill\qed

\medskip 
The following proposition shows that there exists a one-to-one relation between risk aversion processes and investment strategies that depend only on time and the economic state. 
For such investment strategies, the proof shows how the corresponding risk aversion process can be constructed using backward induction.

\begin{prop}\label{PropGamma1}
	Fix a sequence $(\tilde\pi_n)_{0\leq n<T}$, with $\tilde\pi_n:\mathcal{Y}\mapsto\bbr_0^+$. Then, there exists a unique risk aversion process $(\gamma_n)_{n\geq 0}$, with $\gamma_n:\mathcal{Y}\mapsto\bbr_0^+$,  such that	
	$(\tilde\pi_n)_{0\leq n<T}$ is the optimal relative risky asset allocation corresponding to $(\gamma_n)_{0\leq n<T}$.
\end{prop}

\paragraph{Proof of Proposition \ref{PropGamma1}:} We construct the process $(\gamma_n)_{0\leq n<T}$ using backward induction. First, for $n=T-1$,  we have $\gamma_{n}(y)=\tilde\mu(y)/(\tilde\pi_{n}(y)\sigma^2(y))$. For $n<T-1$, consider the function
\begin{align*}
h:\Big(0,\frac{1}{R(y)}\frac{\mu_n^a}{\mu_n^b-(\mu_n^a)^2}\Big)\mapsto (-\tilde\pi_n(y),\infty), \quad
h(x) = \frac{\tilde\mu(y)}{x\sigma^2(y)}\frac{\mu_n^a-R(y)x(\mu_n^b-(\mu_n^a)^2)}{\mu_n^b + \Big(\frac{\tilde\mu(y)}{\sigma(y)}\Big)^2(\mu_n^b- (\mu_n^a)^2)} - \tilde\pi_n(y), 
\end{align*}
where $R(y):=1+r(y)$, $\mu_n^a=\mu_n^a(y)$, and $\mu_n^b=\mu_n^b(y)$. 
The function $h$ is strictly decreasing on its domain, and we set the value of $\gamma_n(y)$ to be the unique root of $h$. {Using Proposition} \ref{PropUhat} and (\ref{u_approx}), {it is simple to check that $(\tilde\pi_n)_{0\leq n<T}$ is the optimal allocation corresponding to the constructed process $(\gamma_n)_{0\leq n<T}$.}
\hfill\qed 

\medskip 

In the following lemma, we compute the Sharpe ratio (\ref{sDef}) for a strategy $\pi$ of the form
\begin{align}\label{piHomog}
\tilde\pi_n = \bar\pi(Y_n).
\end{align}
That is, at any time $n$, in economic state $y\in\mathcal{Y}$, the proportion of wealth invested in the risky asset is given by $\bar\pi(y)$, for some function $\bar\pi:\mathcal{Y}\mapsto\bbr_0^+$. 
For brevity, we write $\bar\pi_y$ for $\bar\pi(y)$. Additionally, we denote the mean $\tilde\mu(y)$ and standard deviation $\sigma(y)$ of excess market returns by $\tilde\mu_y$ and $\sigma_y$, respectively, and the stationary distribution of the Markov chain $(Y_n)_{n\geq 0}$ by $(\lambda_y)_{y\in\mathcal{Y}}$. 
\newpage
\begin{lem}\label{LemSharpe}
	\hfill 
	\begin{itemize} 
		\item[(a)]
	The Sharpe ratio (\ref{sDef}) of the strategy $\pi$ in (\ref{piHomog}) satisfies
	\begin{align}\label{sgen}
	s^{\pi} = \frac{\tilde\mu^{\pi}}{\sigma^{\pi}} = \frac{\sum_{y\in\mathcal{Y}}\lambda_y\tilde\mu_y\bar\pi_y}{\sqrt{\sum_{y\in\mathcal{Y}}\lambda_y\big(\sigma_y^2\bar\pi_y^2 + (\tilde\mu_y\bar\pi_y-\tilde\mu^{\pi})^2\big)}}.
	\end{align}
	Moreover, 
	\begin{align*} 
	(i)\; s^{\pi}\leq \max_{y\in\mathcal{Y}}\frac{\tilde\mu_y}{\sigma_y}, \qquad
	(ii)\; s^{\pi} \longrightarrow \frac{\lambda_{y'}}{\sqrt{\lambda_{y'}(1-\lambda_{y'})^2 + \lambda_{y'}^2\sum_{y\neq y'}\lambda_y}}, \;\text{ if } \tilde\mu_{y'}\to\infty,
	\end{align*} 
	and the maximum in (i) is achieved if and only if $|\mathcal{Y}|=1$. 
	\item[(b)] If $|\mathcal{Y}|=2$ and $\bar\pi_2=\bar\pi_1(1+\delta)$,  formula (\ref{sgen}) can be written as
	\begin{align}\label{sharpe}
	s^{\pi^*}
	&= \frac{1}{\sqrt{\frac{\sigma^2(1)}{\tilde\mu^2(1)}\frac{1  -\lambda_2  + \lambda_2 b^2(1+\delta)^2}{(1 + \lambda_2( a(1+\delta)-1))^2} + \frac{1 -\lambda_2 + \lambda_2 a^2(1+\delta)^2}{(1 + \lambda_2( a(1+\delta)-1))^2}-1}},
	\end{align}
	where $a:=\tilde\mu(2)/\tilde\mu(1)$ and $b:=\tilde\mu(2)/\tilde\mu(1)$. 
	\end{itemize}
\end{lem} 

The previous lemma shows that the Sharpe ratio of $\pi$ is bounded by the largest market Sharpe ratio, and is equal to the market Sharpe ratio in the case of a single economic state. From (i) it also follows that if all states have the same market Sharpe ratio $\theta>0$, 
then the Sharpe ratio of $\pi$ is strictly smaller than $\theta$. 
This can be seen by looking at the {contribution of state $y$ to the variance} 
in the Sharpe ratio formula (\ref{sgen}), which consists of a ``within-state'' variance term, $\sigma_y^2\bar\pi_y^2$, and a ``between-state'' variance term, $(\tilde\mu_y\bar\pi_y-\tilde\mu^{\pi})^2$, where $\tilde\mu^{\pi}=\sum_{y\in\mathcal{Y}}\lambda_y\tilde\mu_y\bar\pi_y$ is the mean excess return of the strategy. Without the between-state term, the Sharpe ratio of $\pi$ would be equal to $\theta$. 

The between-state variance term is also the reason why a larger expected return in a given economic state does not always translate into a higher Sharpe ratio of $\pi$, as can be seen from the limit in (ii). More concretely, if $|\mathcal{Y}|=2$, the Sharpe ratio converges to $\sqrt{\lambda_2/\lambda_1}$ if $\tilde\mu_2\to\infty$. Hence, if $\lambda_2$ is small, i.e., if little time is spent in state $y=2$, then the Sharpe ratio effect of higher expected returns in this state is dominated by the effect of increasing between-state variance. In Lemma \ref{lemDeriv}, we focus on the case of two economic states, and study the sensitivity of the Sharpe ratio of $\pi$ with respect to the model parameters.  

\paragraph{Proof of Lemma \ref{LemSharpe}:} Using that the portfolio return $r_{n,n+1}^{\pi}$ 
is independent of the investment horizon $T$ and that $Y_n$ converges in distribution to a random variable with distribution $(\lambda_y)_{y\in\mathcal{Y}}$, we have
\begin{align*}
\tilde\mu^{\pi} &= \lim_{n\to\infty}\bbe[r_{n,n+1}^{\pi}-r_{n+1}] = \bbe[r_{0,1}^{\pi}-r_1|Y_0\sim\lambda] = \sum_{y\in\mathcal{Y}}\lambda_y\tilde\mu_y\bar\pi_y, \\
(\sigma^{\pi})^2 &= \lim_{n\to\infty}\bbe[(r_{n,n+1}^{\pi}-r_{n+1}-\tilde\mu^{\pi})^2] 
= \bbe[(r_{0,1}^{\pi}-r_{1}-\tilde\mu^{\pi})^2|Y_0\sim\lambda]
= \sum_{y\in\mathcal{Y}}\lambda_y\big(\sigma_y^2\bar\pi_y^2 + (\tilde\mu_y\bar\pi_y-\tilde\mu^{\pi})^2\big).
\end{align*} 
By Jensen's inequality we deduce
\begin{align*}
s^{\pi}
\leq \frac{\sum_{y\in\mathcal{Y}}\lambda_y\tilde\mu_y\bar\pi_y}{\sqrt{\sum_{y\in\mathcal{Y}}\lambda_y\sigma_y^2\bar\pi_y^2}} 
\leq \frac{\sum_{y\in\mathcal{Y}}\lambda_y\tilde\mu_y\bar\pi_y}{\sum_{y\in\mathcal{Y}}\lambda_y\sigma_y\bar\pi_y} 
= \frac{1}{\sum_{y\in\mathcal{Y}}\frac{1}{\tilde\mu_y/\sigma_y}\frac{\lambda_y\tilde\mu_y\bar\pi_y}{\sum_{y'\in\mathcal{Y}}\lambda_y\tilde\mu_{y'}\bar\pi_{y'}}}
\leq \max_{y\in\mathcal{Y}}\frac{\tilde\mu_y}{\sigma_y}.
\end{align*}
If $\tilde\mu_{y'}\to\infty$ for some $y'\in\mathcal{Y}$, then
\begin{align*}
s^{\pi} 
&\longrightarrow \frac{\lambda_{y'}\bar\pi_{y'}}{\sqrt{\lambda_{y'}(\bar\pi_{y'}-\lambda_{y'}\bar\pi_{y'})^2 + \sum_{y\neq y'}\lambda_y\lambda_{y'}^2\bar\pi_{y'}^2}} 
= \frac{\lambda_{y'}}{\sqrt{\lambda_{y'}(1-\lambda_{y'})^2 + \lambda_{y'}^2\sum_{y\neq y'}\lambda_y}}.
\end{align*}
The special case in part (b) follows by simple calculations.
\hfill\qed

\medskip 

\begin{lem}\label{lemDeriv} 
	Let $u:=(\sigma(1)/\tilde\mu(1))^2$. 
	{We have the following characterizations for the sensitivities of the Sharpe ratio $s^{\pi^*}(\delta)$} defined in (\ref{sharpe}):
	\begin{align*}
	\frac{\partial s^{\pi^*}(\delta)}{\partial a}&>0  \quad\Longleftrightarrow\quad  a   < \frac{1+u}{1+\delta} + \frac{\lambda}{1-\lambda}ub^2(1+\delta),  \\
	\frac{\partial s^{\pi^*}(\delta)}{\partial b}&<0 \quad\Longleftrightarrow\quad  b > 0,  \\
	\frac{\partial s^{\pi^*}(\delta)}{\partial \lambda}&>0 \quad\Longleftrightarrow\quad  \lambda > 
	\frac{(a^2+ub^2)(1+\delta)^2-(1+u)(2a(1+\delta)-1)}{(a(1+\delta)-1)((a^2+ub^2)(1+\delta)^2-(1+u))}, \\
	\frac{\partial s^{\pi^*}(\delta)}{\partial \delta} &> 0 \quad\Longleftrightarrow\quad 1+u > a\Big(1+\frac{b^2}{a^2}u\Big)(1+\delta). 
	\end{align*} 
	Furthermore, 
	\begin{align*} 
	\lim_{a\to\infty} s^{\pi^*}(\delta) = \sqrt{\frac{\lambda}{1-\lambda}}, \qquad
	\lim_{b\to\infty} s^{\pi^*}(\delta) = 0, 	\qquad 
	s^{\pi^*}(\delta)\big\lvert_{\lambda=0} = \frac{\tilde\mu(1)}{\sigma(1)}, \qquad s^{\pi^*}(\delta)\big\lvert_{\lambda=1} = \frac{\tilde\mu(2)}{\sigma(2)}.	
	\end{align*} 
\end{lem}

The inequality conditions for the parameters 
in the previous lemma are not restrictive. For instance, let us consider the values of the derivatives in the case $\delta=0$, which corresponds to equal market allocation in both states. First, for the $a$-derivative, we have
\begin{align*}
\left.\frac{\partial s^{\pi^*}(\delta)}{\partial a}\right\lvert_{\delta=0}>0  
&\quad\Longleftrightarrow\quad  a   < 1+u + \frac{ \lambda }{1-\lambda}ub^2 < 1+u, 
\end{align*}
which is trivially satisfied for realistic parameter values.\footnote{For example, for the parameter values in (\ref{musig}) and monthly portfolio rebalancing, $a\approx 2.1$ and $u\approx 66.2$.} 
Next, for the $\lambda$-derivative, we consider two cases:
\begin{align*}
\left.\frac{\partial s^{\pi^*}(\delta)}{\partial \lambda}\right\lvert_{\delta=0}>0 \quad&\Longleftrightarrow\quad  \lambda > \frac{a^2+ub^2-(1+u)(2a-1)}{(a-1)(a^2+ub^2-(1+u))} =\left\{\begin{array}{ll} \frac{1}{a+1},&\; b=a,  \\ \frac{a-(1+u)}{a(a+u)-(1+u)}, &\; b=\sqrt{a}.\end{array}\right.
\end{align*}
First, in the case $b=a$, the market Sharpe ratio is the same in both states. The Sharpe ratio of $\pi^*$ is then maximized for $\lambda\in\{0,1\}$, in which case it is equal to the market Sharpe ratio, while for $0<\lambda<1$ the Sharpe ratio is strictly lower (see discussion following Lemma \ref{LemSharpe}), and it is minimized at $\lambda=1/(a+1)$. 
We also consider the case $b=\sqrt{a}$, and assume $a>1$, leading to a higher market Sharpe ratio in the recessionary state, consistent with empirical evidence. (The case $a<1$ can be similarly analyzed.) The portfolio's  Sharpe ratio is then increasing in $\lambda$ for $0<\lambda<1$, because, as mentioned above, the condition $a<1+u$ is in general satisfied. Note that for the parameter values in (\ref{musig}), $b\approx a^{0.15}<a^{1/2}$, and the portfolio's Sharpe ratio is increasing for $0<\lambda<1$.  

{

{Finally, for the $\delta$-derivative, we assume that $\tilde\mu(1)<\sigma(1)$, which typically holds for stock market returns. For a fixed value of $b>0$, one can then show that the right-hand side of the inequality is decreasing in $a$ for $0<a<a_0(b)$, with $a_0(b):=b\sigma(1)/\tilde\mu(1)$, and is increasing in $a$, for $a>a_0(b)$. The inequality is clearly not satisfied for $a=b$ and is satisfied for $a=a_0(b)$ if and only if}
\begin{align*}
1+u > 2b\sqrt{u}.
\end{align*}
{The above inequality holds for standard parameter values, such as those in} (\ref{musig}) {with monthly portfolio rebalancing. In other words, there exists a threshold $a^*(b)>b$ that $a$ needs to exceed for $s^{\pi^*}(\delta)$ to be increasing at $\delta=0$. Notably, there also exists an upper threshold $a^{**}(b)>a^*(b)$ such that if $a\geq a^{**}(b)$, then the condition is no longer satisfied.}\footnote{For plausible values of the parameters, such as those in (\ref{musig}), the upper bound is very large and not exceeded.} {These thresholds are due to the between-state variance of portfolio returns, and we again refer to the discussion following Lemma} \ref{LemSharpe} for more details. 

}

\paragraph{Proof of Lemma \ref{lemDeriv}:} To show the result for the $a$-derivative, let 
\begin{align*} 
h(a,\lambda,\delta) &:= \frac{(1 -\lambda)(1+u) + \lambda(a^2+ub^2)(1+\delta)^2}{(1 + \lambda( a(1+\delta)-1))^2}
=\frac{(1 -\lambda)(1+u) + \lambda ub^2(1+\delta)^2 + \lambda a^2 (1+\delta)^2}{(1-\lambda + \lambda a(1+\delta))^2} \\
&=:\frac{c_0 + c_1 a^2 }{(d_0 + d_1 a)^2},
\end{align*} 
and note that $s^{\pi^*}(\delta)$ in (\ref{sharpe}) is increasing in $a$ if and only if $h$ is decreasing in $a$. We then have
\begin{align*} 
\frac{\partial h(a,\lambda,\delta)}{\partial a} &= \frac{2c_1a(d_0 + d_1 a)^2-2(d_0 + d_1 a)d_1(c_0 + c_1 a^2)}{(d_0 + d_1 a)^4} = \frac{2c_1a(d_0 + d_1 a)-2d_1(c_0 + c_1 a^2)}{(d_0 + d_1 a)^3},
\end{align*}
and the result is obtained by solving $\frac{\partial h(a,\lambda,\delta)}{\partial a}<0$ for $a$. 
The results for the $\lambda$- and $\delta$-derivatives are obtained in the same way by noting that $s^{\pi^*}(\delta)$ is decreasing in $\lambda$ and $\delta$ if and only if $h$ is decreasing in $\lambda$ and $\delta$.
The rest of the lemma follows directly from the formula for $s^{\pi^*}(\delta)$.
\hfill\qed

\paragraph{Proof of (\ref{pi_concave}):} Let $u:=(\sigma(1)/\tilde\mu^2(1))^2$. The Sharpe ratio formula (\ref{sharpe}) can be written as
\begin{align*} 
s^{\pi^*}(\delta) 
&=\frac{1 + \lambda (a-1)  + \lambda a \delta}{\sqrt{(\lambda+u)(1 - \lambda) + \lambda\big( a^2(1 - \lambda) + u b^2\big)(1+\delta)^2 + 2a\lambda(\lambda-1) (1+\delta) }}
=:\frac{d_0 + d_1 \delta }{\sqrt{c_0  + c_1 (1+\delta) + c_2(1+\delta)^2}}. 
\end{align*}
Let $\kappa(\delta):=\sqrt{c_0  + c_1 (1+\delta) + c_2(1+\delta)^2}$. 
Then simple calculations yield 
\begin{align*} 
\frac{\partial s^{\pi^*}(\delta) }{\partial \delta} 
= \frac{d_1\kappa^2(\delta) - \half(d_0 + d_1 \delta){(c_1 + 2c_2(1+\delta))}}{\kappa^3(\delta)},
\end{align*}
and 
\begin{align*} 
\frac{\partial^2 s^{\pi^*}(\delta) }{\partial \delta^2} 
&= -\frac{(d_0 + d_1 \delta)\big( c_2\kappa^2(\delta)  - \frac{3}{4}(c_1 + 2c_2(1+\delta))^2\big)  + d_1(c_1  + 2c_2(1+\delta))\kappa^2(\delta)}{\kappa^5(\delta)}. 
\end{align*}
For $\delta=0$, we have
\begin{align*} 
\left.\frac{\partial^2 s^{\pi^*}(\delta) }{\partial \delta^2}\right\lvert_{\delta=0} 
&= -\frac{{d_0\big(c_0c_2 - \frac{3}{4}c_1^2 - 2c_1c_2 - 2c_2^2\big) + d_1(c_1  + 2c_2)(c_0  + c_1 + c_2)}}{\kappa^5(0)}\\
&= \Big(\frac{s^{\pi^*}(0)}{d_0}\Big)^{5}\Big(-(a^2u+b^2u^2)\lambda + O(\lambda^2)\Big),
\end{align*}
and the result follows.
\hfill\qed 

\bibliographystyle{plainnat}

\end{document}